\documentclass[aps,prl,superscriptaddress,twocolumn,10pt]{revtex4-1}
\usepackage{times,color,amsmath,amsfonts,amsthm,graphicx,xcolor}
\usepackage[colorlinks = true,breaklinks = true, citecolor = blue]{hyperref}
\usepackage{times,color,amsmath,amsfonts,amsthm,graphicx,xcolor,braket}
\usepackage[version=3]{mhchem}
\usepackage{subfigure}
\usepackage{ulem}
\usepackage[super]{nth}
\newcommand{\cdag}[2] {\hat{c}^{\dagger}_{#1#2}}
\newcommand{\cop}[2] {\hat{c}_{#1#2}}

\newcommand{\NOTE}[1]{ { \color{red}{\textbf{ {[#1]} }} } }

\begin{document}

\title{Constructing Hubbard Models for the Hydrogen Chain using Sliced Basis DMRG}

\author{Randy C. Sawaya}
\affiliation{Department of Physics and Astronomy, University of California, Irvine, CA 92697-4575, USA}

\author{Steven R. White}
\affiliation{Department of Physics and Astronomy, University of California, Irvine, CA 92697-4575, USA}

\begin{abstract}
Sliced-basis DMRG(sb-DMRG) is used to simulate a chain of hydrogen atoms and to construct low-energy effective Hubbard-like models. The downfolding procedure first involves a change of basis to a set of atom-centered Wannier functions constructed from the natural orbitals of the exact DMRG one-particle 
density matrix.  The Wannier function model is then reduced to a fewer-parameter Hubbard-like
model, whose parameters are determined by minimizing the expectation value of the Wannier Hamiltonian in the ground state
of the Hubbard Hamiltonian. This indirect variational procedure not only yields compact and 
simple models for the hydrogen chain, but also allows us to explore the
importance of constraints in the effective Hamiltonian, such as the 
restricting the range of the single-particle hopping and two-particle interactions, and to assess the reliability of more conventional downfolding.
The entanglement entropy for a model's ground state, cut in the middle, is an important property determining the ability of DMRG and tensor networks to simulate the model, and we study its variation with the range of the interactions.  Counterintuitively, we find that  shorter ranged interactions often have larger entanglement.
\end{abstract}

\maketitle

\section{Introduction}

One working definition of a strongly correlated system is one where density
functional theory (DFT) approximations fail, typically because of strong 
entanglement or phases which do not look like Fermi liquids or insulators.  
Traditional approaches to simulating these systems employ DFT in a more limited
way, namely to derive an effective model of the key degrees of 
freedom of the system, which is then simulated with another method that 
can treat strong correlations, such as the density matrix renormalization 
group (DMRG), or quantum Monte Carlo(QMC).  These effective models are typically
constructed by combining the DFT bands into localized Wannier Functions and 
deriving a screened Coulomb interaction from constrained
DFT\cite{cDFT,cDFT_application1,cDFT_application2} or, within a Green's function
framework, the Random Phase Approximation(RPA)\cite{cRPA1,cRPA2,Arya2004}.
In the context of the 
high temperature superconducting cuprates, such methods have been used to 
{\it downfold} to a Hubbard or $t$-$J$ model, with parameters in principle 
coming entirely from the DFT or related methods, but in practice sometimes 
assisted by experiment\cite{Sasha1,Sasha2,Sasha3,Sasha4}.  Although these methods are 
conceptually straightforward and extremely useful, there is generally no path towards systematically improving their accuracy. 

A key limitation of these methods
is that 
they generally require an {\it a priori} parameterization
of an effective Hamiltonian which, in the absence of any rigorous scheme to test different assumptions, provides no guarantee that the excluded terms do not significantly alter the phase diagram of the downfolded model.  There do exist
other techniques based on L\"owdin downfolding\cite{freed1983,Zhou_2010,ten2013} and canonical transformation theory\cite{CT1,CT2,CT3,CT4,CT5} that do not require such parameterizations, but their application to
real materials still remains to be carried out and tested.  In attempting to 
improve downfolding methods in the context of strong correlations, recent work 
has focused on the importance of matching single-body and two-body 
density matrices between the full and effective systems
\cite{Acioli_1994,Zhou_2010,Wagner_2013,Changlani_2013,Changlani_2018,zheng2018}.  
Importantly, it was realized that constructing a low energy Hamiltonian does not require
that one find exact eigenstates of the full system; instead, one can sample from the manifold of low energy states\cite{Changlani_2018,zheng2018}.  This approach allows
one to construct effective models without access to the exact eigenstates of the full
system.

Despite these advances, one thing that 
has been notably missing in these works is the ability to use an accurate 
and reliable strongly correlated method directly on the full system in order to check 
these methods.  The importance of these checks was recognized by Shinaoka, et. al, who
tested an RPA downfolding technique with dynamical mean field theory in the context of mapping multiorbital Hubbard models to low energy models\cite{shinaoka2015}.  The problem has been the lack of methods which can treat strongly correlated real materials directly, with clear-cut control of the accuracy. This situation has
changed in the last few years, however, as there has been considerable improvement in adapting
a number
of strongly correlated methods to work directly on solids\cite{Motta_2017,Motta_2020}. Progress has
been made by focusing 
attention on some of the simplest systems which still have key features of a solid, such as  a chain of equally spaced 
hydrogen atoms. This system is a natural playground to study Mott physics in
one dimension, with a one dimensional Hubbard model a natural candidate for describing the low energy space. The hydrogen chain
was the subject 
of a large multi-method study\cite{Motta_2017,Motta_2020}, where it was found
that a number
of methods can treat this system very accurately, including sliced basis DMRG(sb-DMRG)\cite{Stoudenmire_2017}, the technique used here.

The progress in simulations allows one not only to test weak-coupling-based
downfolding methods, but also to develop new downfolding schemes which use  the strong coupling algorithms directly.  Here, we utilize very accurate sbDMRG
calculations on hydrogen chains of sizes up to 100 atoms to test downfolding ideas. We couple this with DMRG calculations on the low energy models, which range from simple Hubbard models to Hubbard-like models with extended hopping and density-density interactions. 
Instead of constructing Wannier functions based  on independent particle
bands, we construct ``natural Wannier functions"
\cite{Koch_Goedecker_2001} from the natural orbitals of
the fully-interacting single-particle density matrix.  This allows the technology developed for ordinary Wannier
functions to be transferred over to the fully interacting regime.  Within this framework, we
can consider what terms are needed (including the range of both 
hopping and Coulomb interactions) by generalizing the functional minimization of the
Peierls-Feynman-Bogoliubov variational principle used 
in \cite{Schuler_2018} to both the single- and two-particle interactions.  Moreover,
we can evaluate the 
dependency of the needed coupling on the range of excited states one wants to 
represent with the effective model.

DMRG and tensor network methods are based on low entanglement.  For DMRG, what matters is the total entanglement when the system is cut in two. Two or three  dimensional systems have entanglement entropies which grow proportionally with the transverse area.  In DMRG, in 2 or 3D, the system is mapped onto a snake-like path and the higher entanglement is tied to longer-range couplings along this path.  Therefore, it is natural to expect that even in a different context, namely a 1D system with  decaying long-ranged interactions, longer ranges would correspond to higher entanglement. We test this proposition and find that it is generally untrue.  In the context of the Hubbard-like models considered here, shorter-ranged models tend to have slightly more entanglement than longer-ranged models. Nonsmoothness in the interaction also tends to increase increase entanglement.

In the following section, we introduce the hydrogen chain and the corresponding
numerical methods used to solve the system and downfold it to an effective 
model.  Then in section \ref{sec:EffH}, we explore different types of effective 
models that we find can be used to describe the hydrogen chain as well as
investigate the necessity of various terms in the model Hamiltonian. We also study the
entanglement dependence on the range of interactions. Finally, 
we conclude in section \ref{sec:conc}.

\section{Hydrogen Chain and Numerical Methods}

In this section, we give a brief introduction to the Hydrogen chain and sb-DMRG and introduce our method for constructing Wannier 
functions.

\subsection{Hydrogen Chain}

As used in this work, the hydrogen chain (H-chain) consists of $N$ protons and electrons.  The protons are held fixed and 
are equally spaced by a distance $R$. Hydrogen chains would be unstable chemically, but for electronic structure calculations they are realistic in the sense of having three dimensional electron wavefunctions governed by the Schr\"odinger equation, with long range Coulomb interactions.
Most importantly, it can be regarded as a realistic model of a 1D solid, with an interesting phase diagram as one varies $R$.
The H-chain is also simple enough to be accessible to modern many-body methods\cite{Motta_2017,Motta_2020,HChain1,HChain2,HChain3,HChain4}, including DMRG methods. 
The H-chain is also thought to be equivalent to  the Hubbard model in the large $R$ limit.
These features, together with it's computational tractability, make the H-chain an excellent candidate to study the progression from a real system to an effective model.

\paragraph{\textbf{Electronic Correlations}}

The correlations within the H-chain are controlled by a single parameter, $R$, 
the atomic spacing.  In a localized atomic orbital picture, $R$ controls the degree of overlap between neighboring orbitals which in turn determines the hopping strength.  
At small $R$, the orbitals heavily overlap leading to more hopping which, in the 
extreme limit, begins to resemble a 1D electron gas.  As $R$ is increased, the lack of 
overlap between neighboring atoms suppresses the hopping, leading to a more insulating
state. This is the standard scenario for a Mott metal-insulator transition,
although the 1D nature of the chain can alter the expected physics. 

For $R > 1.8a_B$, the low-energy physics is dominated by a single band composed primarily of the atomic 1s orbitals, with higher orbitals mixing in slightly.  At this stage, increasing $R$  increases
the electronic correlations, roughly corresponding to increasing $U/t$ in the Hubbard model by decreasing $t$.  At a critical $R_c < 1.8a_B$, instead of following the single band Hubbard model for smaller $U/t$, additional bands start to become occupied.  These additional bands are diffuse and emerge from the $2s$, $2p_x$, and $2p_y$ atomic orbitals.
The occupancy of the 1s band drops below half-filling, and metallic 
behavior is observed\cite{Motta_2020}.  In this more complicated regime, the system would require a multi-band
effective model.  We hope that the techniques developed in this paper will be useful in constructing multiband models, but for simplicity we restrict ourselves here to the single-band regime, with $R \ge 1.8a_B$. 

\paragraph{\textbf{Boundary Conditions}}

In a real hydrogen chain, the tail of the electronic wavefunction extends a 
moderate distance beyond the nuclei that make up the chain.  When dealing with a 
finite chain as a segment of an infinite chain, however, the electrons can no longer
spread beyond the finite segment, since the other atoms that make up the larger, infinite
chain obstruct them from doing so.  Without taking this effect into account, the end
spill-off of the electrons will result in a reduction of the bulk electron density, 
slowing the convergence to the thermodynamic limit.  Although this effect is very minor
at large $R$, it becomes  important for sufficiently small $R$, and results in a reduction in $R_c$ for shorter chains\cite{Motta_2020}.
If periodic boundary conditions are used, this boundary effect does not occur, but periodic boundaries slow the convergence of DMRG.  Instead, we use hard walls
placed at a distance $R/2$ beyond the first and last atoms in order to mimic the presence
of other atoms that would otherwise be there in the infinite chain.  While there are still some edge effects, we find that they are much smaller
than with open ends. 

\subsection{sb-DMRG Calculations}

Adapting DMRG to real systems requires parameterizng the continuous degrees of freedom 
into localized ``sites" with a locally defined Hilbert space.  In the conventional approach the sites are linear combinations of Gaussian basis functions. A severe drawback of this approach is the large number of two-electron terms in the resulting Hamiltonian, scaling as $N_b^4$, where $N_b$ is the number of basis functions. If one instead used a real-space grid representation, with $N_g$ being the number of grid points, one gets $N_g^2$ interaction terms but $N_g \gg N_b$.  Sliced bases are a compromise, with favorable scaling and an intermediate number of sites\cite{Stoudenmire_2017}.

In sb-DMRG we use a real-space grid along one direction and a localized set of 2D Gaussian basis functions spanning the other two transverse 
directions, defining a ``slice" at each grid point.  This is particularly advantageous when dealing with systems
extending much more in one direction than in the transverse direction, such as chains of atoms. The entanglement of the system in the sliced basis is favorable, not much above that of the corresponding effective Hubbard model. In contrast, sets of conventional basis functions must be localized before being used in DMRG, since extended molecular orbitals can give rise to volume-law entanglement scaling, which becomes dominant on long chains.

In addition to maintaining an
area-law entanglement scaling, restricting these functions to slices makes each function orthogonal to all
other functions not on the same slice, reducing the number of interaction terms from quartic to quadratic 
in the number of grid points.  The resulting long-range Coulomb interactions may then by 
compressed via an SVD scheme to produce an MPO with a modest bond dimension which is nearly
independent of system size\cite{ChanWhite2016,Zaletel2020}.  This technique allows the method to scale essentially linearly 
with the number of atoms, making large chains of atoms accessible to DMRG-level accuracy.

\paragraph{\textbf{Constructing the sliced-basis Functions}} \label{para:basis}

The  2D transverse functions defining the slices are taken from standard 3D Gaussian basis set functions  defined by 
four numbers, $(\zeta,n,m,k)$ with each function taking the form,
\begin{align}
g_i(\vec{r}) = x^n y^m z^k e^{-\zeta \vec{r}^2}
\end{align}
The values of $n$, $m$ and $k$ determine the character of the function(i.e. $s$-type, $p$-type etc.) and $\zeta$ controls the width.
Some of these primitive Gaussians (especially the s functions) are contracted with others of the same type to form new contracted basis functions. In the case of uncontracted Gaussians, the $z^k$ term is removed and the resulting 2D Gaussian is used for each slice. In the case of contractions, the $z^k$ comes in making the 2D contraction coefficients from the 3D contractions; thus, a sharp localized Gaussian gets a bigger weight for slices near the nuclei.   In addition, we may use a two-step calculation, where an initial sb-DMRG is used to get a local single particle density matrix for each slice, which is then used
to contract the slice functions further, using slice natural orbitals. 
Since the Wannier functions used for downfolding the hydrogen chain come from the 1s band, for most of the calculations we start with only $S$ ($n=0, m=0, k=0$) Gaussians (taken from a CC-pVDZ basis) and then use the slice natural orbitals to reduce to a single function per slice.  In this case with only one function per slice, the Hamiltonian takes the simple form,
\begin{align}
	\hat{H}_{\text{sb}} =\ \sum_{\sigma,nn'} t^\text{sb}_{nn'} \cdag{\sigma}{n} \cop{\sigma}{n'}\ + 
			\frac{1}{2} \sum_{\sigma,nn'} V^\text{sb}_{nn'} \hat{n}_{\sigma n} \hat{n}_{\sigma' n'}
	\label{eqn:H1}
\end{align}
where $n$ and $n'$ enumerate the slices and sb indicates sliced basis.

\begin{figure}
\centering
\includegraphics[width=\columnwidth]{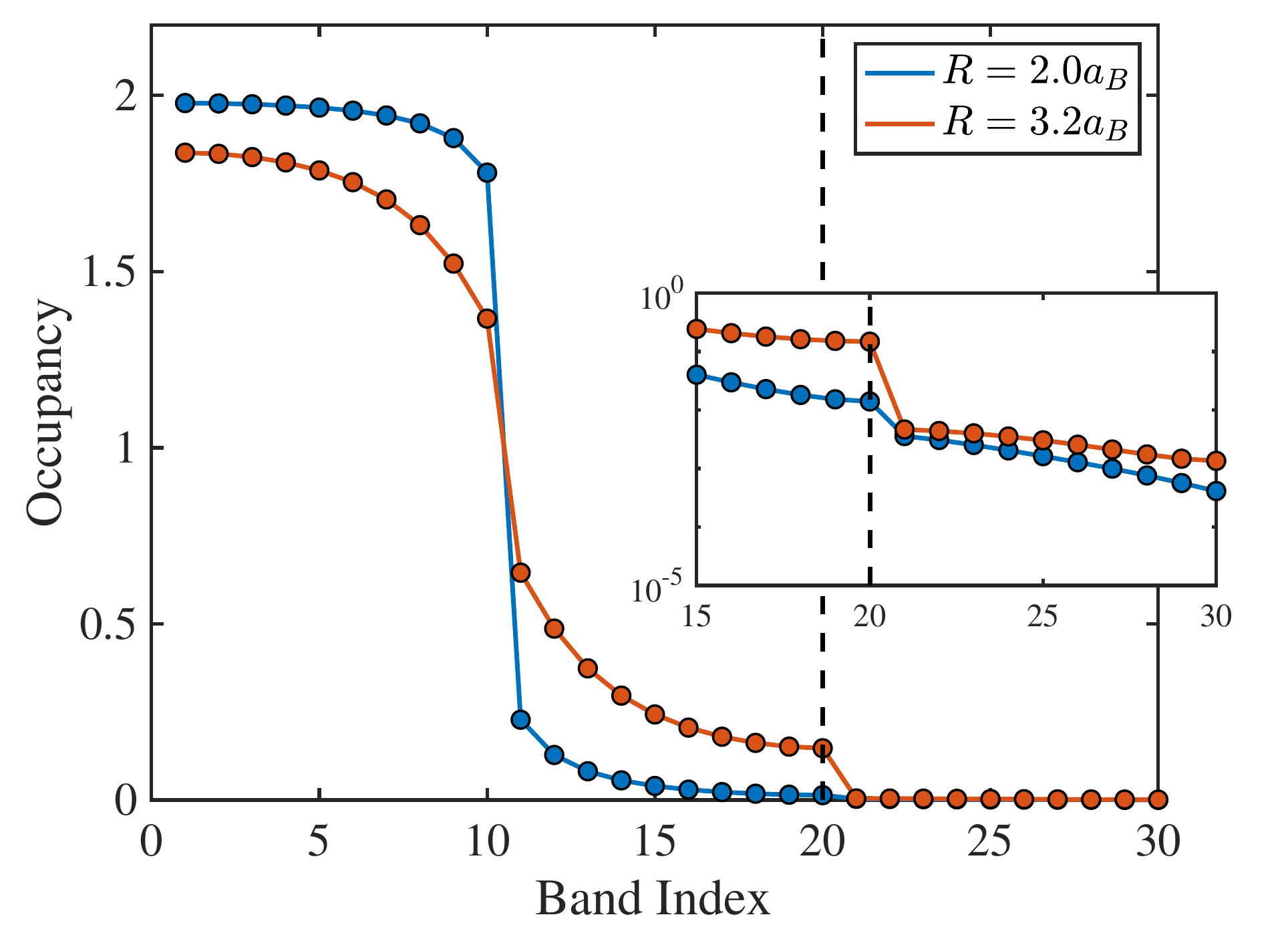}
\caption{Eigenvalues of the single-particle correlation matrix from the ground state
of H20 giving the natural orbital occupancy.  The first 10 states are doubly occupied and represent Hartree-Fock type states.  The second 10 states have
small but non-negligible occupancy.  States beyond the first 20 show occupancy on the order of $10^{-3}$ and are truncated to give a 
downfolding approximation.}
\label{fig:occ}
\end{figure}

\subsection{Downfolding to an Effective Model}

\paragraph{\textbf{Constructing the Wannier Functions}}

Wannier functions(WF) are linear combinations of Bloch orbitals, which are eigenstates 
of a  one-body Hamiltonian.  The one-body Hamiltonian is the result of a
specific approximation, such as a tight-binding model or density functional calculation. Even within these approximate theories, where 
the electronic ground state is fully specified by the single-particle states, the inherent gauge freedom 
of the Bloch orbitals makes the resulting WFs  non-unique.  It wasn't until
the introduction of the ``maximally localized" criterion\cite{marzari1997} that WFs began 
taking on a wide variety of tasks in e.g. linear-scaling algorithms or 
electronic structure calculations (see \cite{MLWF} for
a nice review).

What about constructing WFs from a fully interacting approach, such as DMRG, where 
the single-particle Bloch orbitals that make up the WFs
are no longer well
defined?  As pointed out  in \cite{Koch_Goedecker_2001},  one can
use the single-particle density matrix, which is well-defined even in
an exact theory,
as a surrogate for the Hamiltonian.  One then takes its eigenstates (the natural orbitals), sorts them into bands, and constructs WFs.  Of course,
the full interactions fractionally populate the omitted bands and reducing
to the WFs is an approximation.  A good measure of the size of the error in
going to the WF basis is in the total occupancy of all the omitted natural orbitals which we find to vary between $10^{-3}$ to $10^{-4}$ depending on $R$(see Fig. \ref{fig:occ}), making this a good approximation for the systems studied here.

This
is the basic approach used here with sb-DMRG, but
we can 
go even further and use eigenvectors of not just the ground state 
single-particle density matrix, but rather a sum of density matrices from a set of low-lying many-particle excited states in addition to the ground 
state, in order to capture information that describes the physics beyond 
the ground state.  Generally, adding density matrices results in new
eigenstates which attempt to cover the space of all the important states of the individual density matrices. We find that using this sum with a well-chosen set of excited states leads to a better 
representation of the low-energy physics in the downfolded model, such
as the gaps between low-lying states.
Choosing which low-lying states to incorporate into the sum of density matrices is key since too many incorporated states
will lead to more truncation unless one uses a less simplified model.  Too few may contribute to errors in reproducing the low-energy physics.  

From (\ref{eqn:H1}), we can see that expressing the original 3D H-chain Hamiltonian in the sliced-basis has resulted in a 
1D single-band lattice Hamiltonian.  Reduced to 1D, we can now regard this system from a Luttinger Liquid perspective, 
for which all low-lying excitations are bosonic spin and charge excitations.
However, extra electrons added to the H-chain generally occupy additional bands which tend to be diffuse\cite{Motta_2020}. A single-band effective model 
cannot properly capture such excitations, so we explicitly omit charge excitations from consideration in building our effective models. The charge excitations in the resulting models are expected to be qualitatively different from charge excitations of the H-chain.  In
contrast, the spin excitations do primarily live in the original single band.  Therefore, in addition to the ground state 
density matrix, density matrices from states with one- and two-spinon pair excitations are added, i.e. total spin $S^z_{\rm total}\le 2$.  Since the excitations only affect the spin-independent Wannier functions, we only create spin excitations in the spin-$z$ direction, speeding the calculations.

Specifically, the single-particle density matrix is calculated using the ground state $\ket{\Psi_0}$, $1$-spinon pair excited state 
$\ket{\Psi_1}$ and $2$-spinon pair excited state $\ket{\Psi_2}$ of the H-chain as,
\begin{align}
	C_{ij}^{(0)} = \bra{\Psi_0} \cdag{i}{\uparrow} \cop{j}{\uparrow} + \cdag{i}{\downarrow} \cop{j}{\downarrow} \ket{\Psi_0} \nonumber \\
	C_{ij}^{(1)} = \bra{\Psi_1} \cdag{i}{\uparrow} \cop{j}{\uparrow} + \cdag{i}{\downarrow} \cop{j}{\downarrow} \ket{\Psi_1} \\
	C_{ij}^{(2)} = \bra{\Psi_2} \cdag{i}{\uparrow} \cop{j}{\uparrow} + \cdag{i}{\downarrow} \cop{j}{\downarrow} \ket{\Psi_2} \nonumber
\end{align}
The mixed density matrix is then given by $C_{ij} = C_{ij}^{(0)} + C_{ij}^{(1)} + C_{ij}^{(2)}$ and the natural orbitals,
$\{\xi_k\}$, are readily obtained by diagonalization.  

For an $N_a$-atom hydrogen chain, the $N_a$ most occupied natural orbitals are
retained whereas the rest are projected out.
In the weakly correlated regime, the spectrum of the ground state density matrix exhibits a sharp cutoff between occupied and unoccupied states whereas in the strongly correlated regime, the spectrum is more flat, reflecting the tendency for these states 
to mix.  However in both of these regimes, there exists a sharp cutoff between the first $N_a$ states and the rest, as shown in 
Fig. \ref{fig:occ}.  Spectra from density matrices of low-lying states(e.g. those generated by spinon excitations) also 
demonstrate the same feature, indicating that the ground state and a group of low-lying excited states are confined 
to the states below this cutoff.  Therefore by diagonalizing a mixture of density matrices, information regarding 
excited states can be captured explicitly and used to construct the effective model.  The $N_a$ remaining natural orbitals are then localized around each atom by projecting
the 1D position operator into this basis and re-diagonalizing.  This is exactly analogous to the spread minimization 
procedure used in the construction of maximally localized WFs.
In 3D systems, the localization functional must be variationally minimized, but reduces in 1D to
diagonalizing the projected position operator\cite{MLWF}.

\begin{figure}
\centering
\includegraphics[width=\columnwidth]{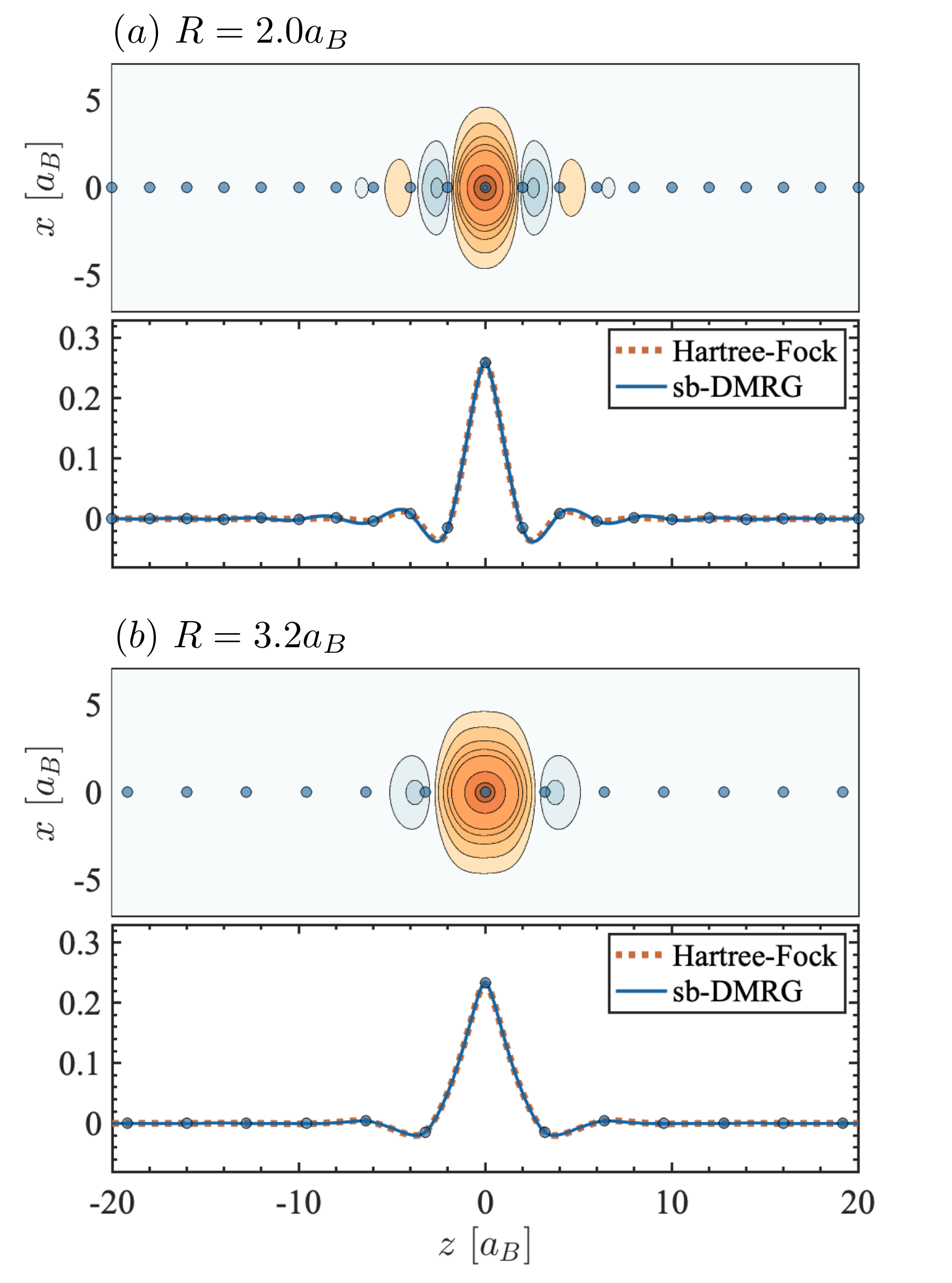}
\caption{Translationally invariant Wannier function for $(a)\ R=2.0a_B$ and $(b)\ R=3.2a_B$ constructed from the single-particle
density matrices of H100.
The top panels in each subplot show the full 3D Wannier functions in real-space integrated over $y$ and the lower panels, 
the 1D representation within the sliced basis as a function of $z$ for both sb-DMRG and Hartree-Fock derived functions.}
\label{fig:WFs}
\end{figure}


The procedure described thus far yields slightly different WFs describing each atom.  This is a consequence 
of using open boundary conditions and as such, only the WFs near the edge differ much from 
those in the bulk.  In order to better reproduce the model in the thermodynamic limit, we restore translational invariance 
among the WFs by first aligning them so that all the maxima coincide at one point then averaging point-wise across all functions, weighing each 
function by their proximity to the center of the chain.  The translationally invariant WF shown in Fig. \ref{fig:WFs} is then redistributed 
to each atom.

\begin{figure}
\centering
\includegraphics[width=\columnwidth]{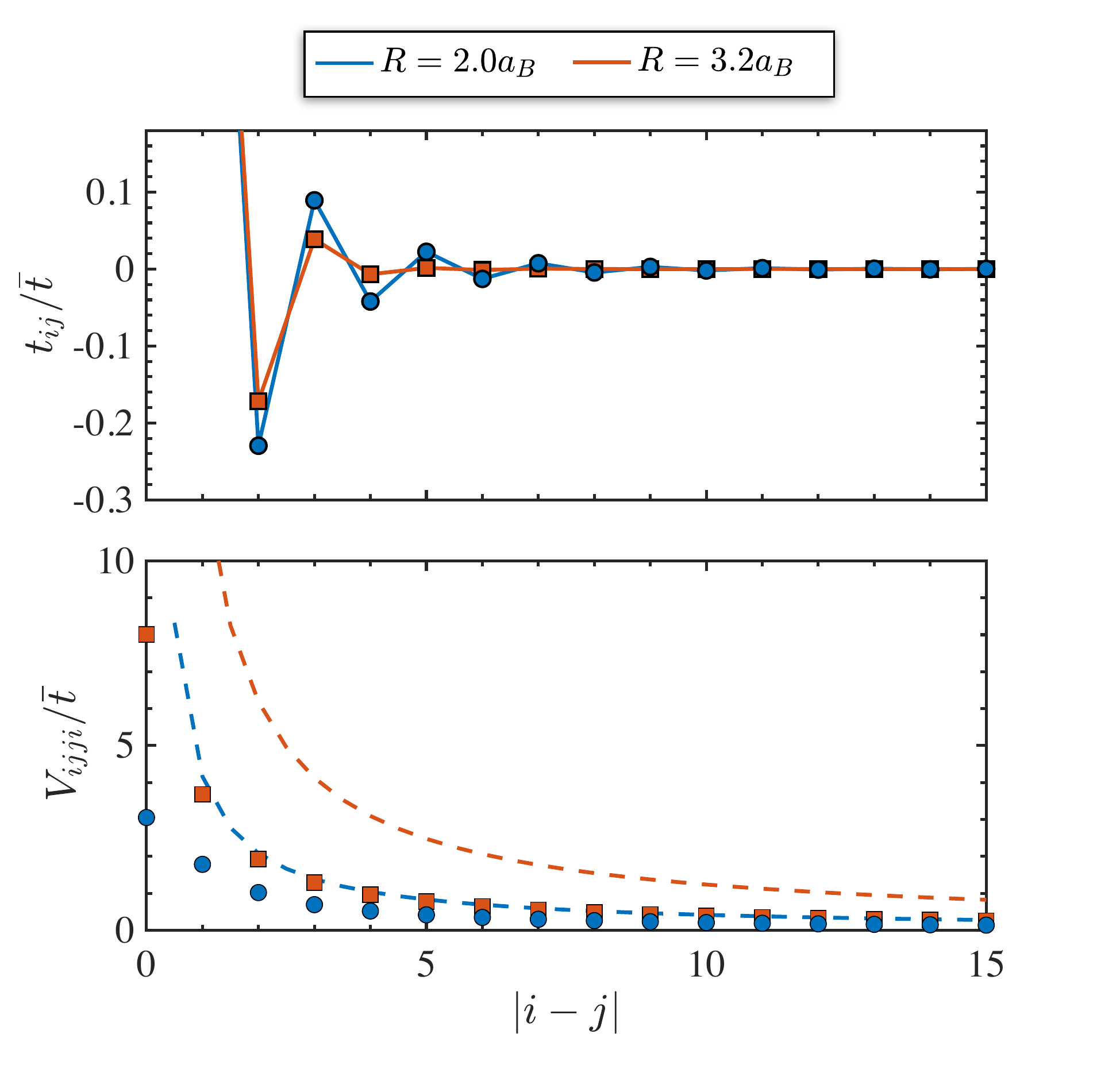}
\caption{Hamiltonian parameters from transforming the H-chain Hamiltonian
into an effective model using the Wannier functions for H100.  All parameters are scaled by the average value of the nearest neighbor 
hopping denoted by $\bar{t}$.  The top panel shows the single-body terms and the lower panel, the two-body interaction terms.
The dashed lines in the lower panel are $1/r$ lines in units of $\bar{t}$.}
\label{fig:WFHparams}
\end{figure}

The WFs are defined by the coefficients $w_{n i}$, where $n = 1,2,...,N_z$ enumerates the grid and $i = 1,2,...,N_a$, 
the functions.
The Hamiltonian then takes the form,
\begin{align}
	\hat{H}_{\text{WF}} =\ \sum_{\sigma, ij} t^\text{WF}_{ij} &\cdag{\sigma}{i} \cop{\sigma}{j}\ +
			\frac{1}{2} \sum_{\sigma, ijkl} V^\text{WF}_{ijkl} \cdag{\sigma}{i} \cdag{\sigma'}{j} \cop{\sigma'}{k} \cop{\sigma}{l} 
			\nonumber \\
	&t^\text{WF}_{ij} = \sum_{nn'} w^*_{i n} t^\text{sb}_{nn'} w_{n' j} \\ 
	V^\text{WF}_{ijkl} &= \sum_{nn'} w^*_{i n} w^*_{j n'} V^\text{sb}_{nn'} w_{k n'} w_{l n} \nonumber .
	\label{eqn:Heff}
\end{align}
The resulting Hamiltonian parameters are shown in Fig. \ref{fig:WFHparams}.

As previously mentioned, in typical applications, Wannier functions are constructed
from single-particle orbitals of mean-field calculations, so it is natural to 
compare the performance of such functions to those obtained from sb-DMRG.
Although density functional approaches may be better, for simplicity we compare unrestricted Hartree-Fock  in place of sb-DMRG to construct these functions and as Fig. \ref{fig:WFs} indicates, 
we find very similar Wannier functions. In this case, we see that a conventional Wannier construction would likely be satisfactory, although we expect higher accuracy from our sb-DMRG approach.

\paragraph{\textbf{Truncating Interaction Terms}}

The number of interaction terms in the Wannier effective Hamiltonian scales as $N^4$, which is inconvenient and costly.  How can one renormalize down to fewer terms?
One approach is to perform unitary transformations in the many-particle space which can be constructed to remove off-diagonal interactions, at the expense of creating new interactions involving more than two particles\cite{CT1,CT2,CT3,CT4,CT5,Kirtman1981,Hoffman1988}.  However, these canonical transformation do not reduce diagonal interactions, say of the form $n_i n_j$ for distant sites $i$ and $j$.  Thus,  we could not use these to form a local Hubbard-like model.

Another, completely
different approach starts by asking: is there another,
simpler Hamiltonian, which has approximately the same
ground state?  For example, in an insulator, the precise form of the long-range Coulomb interaction likely does not matter in the determination of the ground state, as long as the difference is smooth, and the same change is made in the nuclear-electron potential.
A practical way to optimize such effective Hamiltonians is to solve the effective model (say, with DMRG, for each set of parameters) and then minimize over its parameters the expectation value of the original Hamiltonian within the model's ground state.  If one approaches the ground state energy of the original Hamiltonian, then the effective Hamiltonian is doing its job. If, in addition, a set of low energy eigenstates also matched, one would have an excellent effective Hamiltonian. Clearly, one could need a different Hamiltonian depending on the excitations being considered.  Charge excitations would need to have the long-range Coulomb terms kept.  The ground state energy of the effective Hamiltonian could be adjusted to be correct simply by adding a constant.  Other terms might need to be added and optimized for, to get excited state energies and wavefunctions correct. 

Let the effective Hamiltonian we want to determine take the form
\begin{equation}
	\hat{H}_{\text{eff}}[t_{ij}, V_{ij}] = \ \sum_{ij} t_{ij} \cdag{\sigma}{i} \cop{\sigma}{j}\ +
			\frac{1}{2} \sum_{ij} V_{ij} \hat{n}_i \hat{n}_j
			\ \ \text{.}
	\label{eqn:Heff}
\end{equation}
Here we have chosen a simpler density-density two-electron interaction.  We will also restrict the range of $t_{ij}$ and $V_{ij}$ in order to get a simple effective model, so many of these
coefficients will be zero.

Taking the ground state of $\hat{H}_\text{eff}$ to be $\ket{\Psi_\text{eff}} = \ket{\Psi_\text{eff}[t_{ij}, V_{ij}]}$, we choose the optimal elements of $t$ and $V$ by minimizing
\begin{align}
	\Phi[t_{ij}, V_{ij}] = \bra{\Psi_\text{eff}}  \hat{H}_\text{WF}  \ket{\Psi_\text{eff}} \ \ \text{.}
	\label{eqn:phi}
\end{align}
This procedure is used in \cite{Schuler_2018} for the case of reducing long-range two-particle interactions to an on-site interaction.  In
that work, only the interaction terms were adjusted whereas in the
present case, the method has been generalized to adjust both the
one-particle and two-particle terms.  

\begin{figure*}
\centering
\includegraphics[scale=0.5]{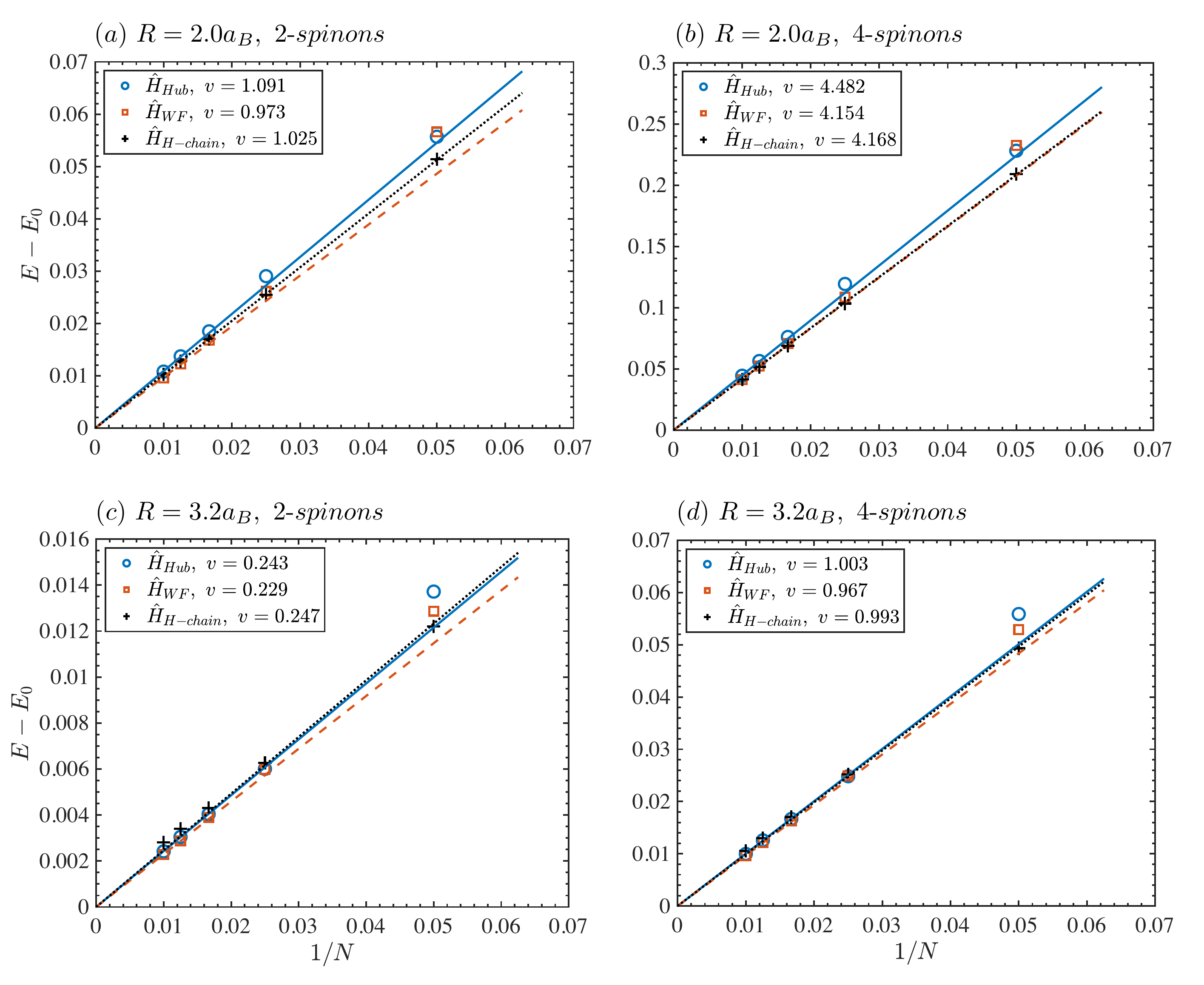}
\caption{Spin velocities for the H-chain and two effective models defined in (\ref{eqn:HWF}) and (\ref{eqn:Hhub}).  The top row shows $(a)$ 2-spinon and $(b)$ 4-spinon excitation 
gaps at $R=2.0a_B$ for the H-chain, its corresponding directly transformed model denoted as $\hat{H}_\text{WF}$ and optimal Hubbard model, $\hat{H}_\text{Hub}$.
The same is done at $R=3.2a_B$ in the lower row for the $(c)$ 2-spinon and $(d)$ 4-spinon excitations.  The numbers
in the legend correspond to the slope of the fitted lines which give the spin velocities.}
\label{fig:gaps}
\end{figure*}

At each step of the optimization, the function evaluation is carried out by first defining a Hamiltonian according to the variational parameters $t_{ij}$ and $V_{ij}$ and then solving for the ground state using DMRG, yielding $\ket{\Psi_\text{eff}}$.  Then (\ref{eqn:phi}) can be used to evaluate $\Phi[t_{ij}, V_{ij}]$
by taking the energy overlap with respect to this wavefunction.
The number of sweeps for each DMRG run is very small---usually no more
than one or two---since the wavefunction from the previous step can be
used as a starting point for the DMRG
at the current step.  

In some cases, we found that this optimization could get stuck.  As a simple fix, instead of optimizing all the parameters at once, we first optimize over the elements of $V$, keeping $t$
fixed, then over the elements of $t$, keeping $V$ fixed, then over $V$ again, etc.

\paragraph{\textbf{Truncating from $V_{ijkl} \rightarrow V_{ij}$}}

The presence of four-index terms, $V^\text{WF}_{ijkl}$, greatly increases the MPO bond dimension, strongly restricting the length of the chains we could study.
Inspecting the four-index terms, we find that the majority of elements with $i \neq l$ and $j \neq k$ are small, on the order of $10^{-6}$ -- $10^{-5}$, indicating a truncation of these terms may have negligible effects on the final model.  Therefore in order to treat much larger system sizes, we first reduce to an intermediate model, keeping the hopping fixed, but reducing $V^\text{WF}_{ijkl}$ to the diagonal $V_{ij}$ form shown in Eq. (\ref{eqn:Heff}), without restricting the range of the interactions.  The diagonal form is approximated from the four-index terms as
$V_{ij} \approx V^\text{WF}_{ijji}$.  To establish that this is a good approximation, we first assume
\begin{align}
	V_{ij} = V^\text{WF}_{ijji} + \delta_{ij} \epsilon_i
\end{align}
and perform the optimization described in the previous section over the set $\{\epsilon_i\}$ on medium length chains, up to $N_a = 20$ atoms.  On such systems, the corrections turn out to be quite small, $\{\epsilon_i\} \sim 10^{-2}$.  When comparing the ground states energies and wavefunctions between the original system and it's diagonal approximation, we find errors on the order of $10^{-3}$ for both the energy difference and wavefunction overlap, indicating that the truncation of these terms is justifiable.  Under this approximation, the Hamiltonian becomes, 
\begin{align}
	\hat{H}_{WF} = \sum_{\sigma,ij} t^\text{WF}_{ij} \cdag{\sigma}{i} \cop{\sigma}{j}\ + 
			\frac{1}{2} \sum_{\sigma,ij} V^\text{WF}_{ijji} \hat{n}_{\sigma i} \hat{n}_{\sigma' j}
	\label{eqn:HWF}
\end{align}
We will use this approximation for $\hat{H}_{WF}$ for the rest of the paper.

\section{Exploring Different Effective Models}\label{sec:EffH}

This section describes various effective models and includes a discussion on the range of one- and two-body terms required to achieve an accurate effective model.  

\subsection{\textbf{Short-range Models}}

One fundamental question regarding effective models is the necessity 
of the various terms that make up the Hamiltonian such as the next-nearest neighbor hopping, third-nearest neighbor hopping and the range of the Coulomb interaction.  Given the method outlined in 
the previous section, we can apply it to a host of effective models, each with a different 
set of single-particle and two-particle terms.  Then by comparing each of the resulting models to the original, we can gauge the importance of
the hopping and the two-body interactions, specifically their relative magnitudes and range.

In what follows, we first establish a baseline effective model using 
(\ref{eqn:HWF}), which includes the full-range hopping and two-particle interactions.  We can then restrict the range of each of these terms,
and optimize over the remaining terms.  Finally, we can reduce the effective model to a pure Hubbard model defined by an optimal hopping, $t$ and optimal on-site interaction $U$.

\begin{figure}
\centering
\includegraphics[width=\columnwidth]{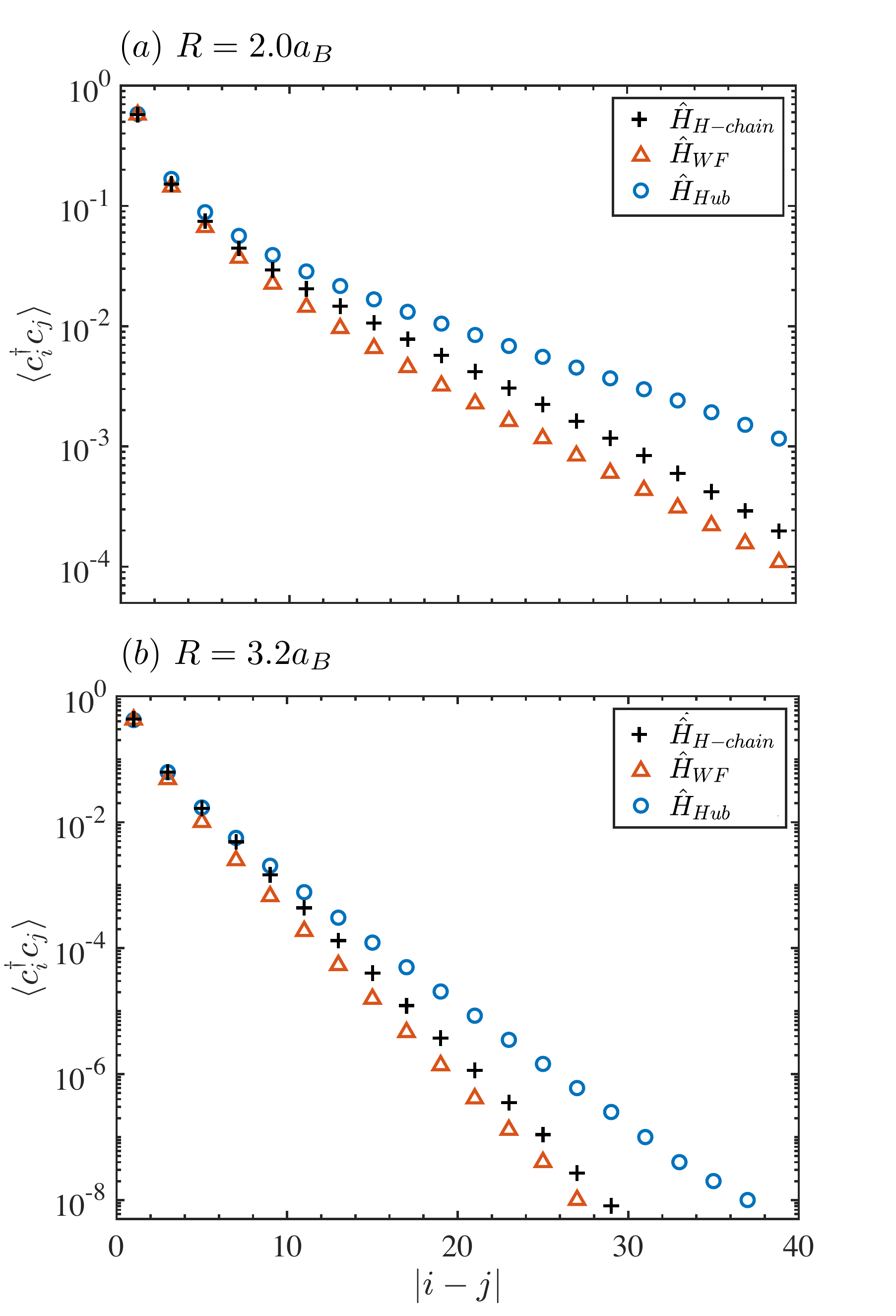}
\caption{Single-particle correlations for the H-chain, its 
corresponding directly transformed model defined in (\ref{eqn:HWF}) and optimal Hubbard model defined in (\ref{eqn:Hhub}) for
H100 at $(a)$ $R=2.0a_B$ and $(b)$ $R=3.2a_B$.  The correlators are calculated from the middle of the 
chain to the right edge as defined in (\ref{eqn:gr}).  Note here that, the WFs are used to transform the H-chain correlation functions into the same space as the effective models.}
\label{fig:cors}
\end{figure}

\paragraph{\textbf{Using $\hat{H}_{WF}$ as an Effective Model}}

A natural baseline can be established by taking the effective model to be $\hat{H}_{WF}$.  This is the Hamiltonian that
resulted from directly transforming the H-chain Hamiltonian with the WFs and subsequently replacing
$V^\text{WF}_{ijkl} \rightarrow V_{ij}$ as was done in Eq. (\ref{eqn:HWF}). 
The accuracy of this and other effective models can be tested by comparing
several quantities such as the spin velocity, single-particle Green's 
function and spin-spin correlation to those of the original H-chain.  These
are meant
to quantify both the ground state properties as well as properties of the low-lying excited states.  Since at this stage, no Hamiltonian parameters have been optimized, any errors
in $\hat{H}_{WF}$ can only result from either the truncation of the 
natural orbitals or the truncation of $V^\text{WF}_{ijkl}$.  

The spin velocity is derived by measuring how the energy gap
between the ground state and a spin-excited state scales with system size.  This can 
be done for $1$ and $2$ pairs of spinons to get information on how well the effective 
model reproduces the low-energy physics of the original system. 
This is shown in Fig. \ref{fig:gaps} for several effective models as
well as the original H-chain.  For smaller systems of $20$ and
$40$ atoms, 
the finite size effects can cause large errors in the gaps, both at
$R=2.0a_B$ and $R=3.2a_B$.  At larger system sizes, these effects
become much more negligible and it can clearly be seen that the 
gaps 
of the effective models begin to closely approximate the gaps
in the original H-chain.  Comparing the actual spin velocities, we find differences
less than $10\%$ between the H-chain and $\hat{H}_\text{WF}$.  

\begin{figure}
\centering
\includegraphics[scale=0.45]{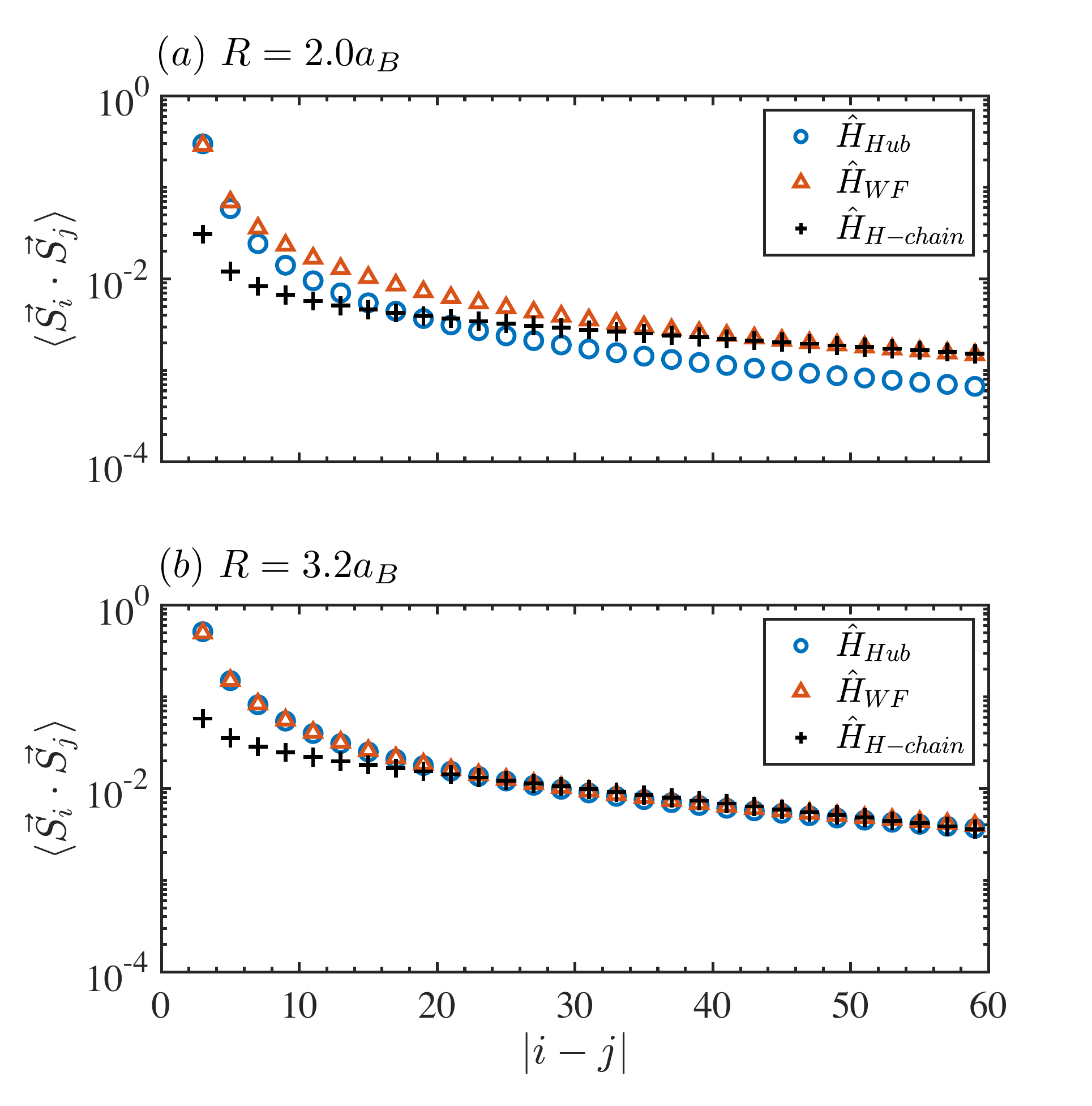}
\caption{Spin-spin correlations at $(a)$ $R=2.0a_B$ and $(b)$ $R=3.2a_B$ for the original H-chain and its corresponding effective models defined in (\ref{eqn:HWF}) and (\ref{eqn:Hhub}).  The correlators are calculated from the left-most edge of the 
chain to the midpoint for a system of 100 atoms.  The H-chain correlations have been scaled such that the point at site 50 match that of $\hat{H}_\text{WF}$. }
\label{fig:sscors}
\end{figure}

In addition to the spin velocity, the single-particle correlations
can be computed and compared across models.  These are defined here as
\begin{align}
	g_i =  \braket{ \cdag{N/2}{\uparrow} \cop{i}{\uparrow} + \cdag{N/2}{\downarrow} \cop{i}{\downarrow} }
	\label{eqn:gr}
\end{align}
and are expected to decay as a power law(up to log corrections) for gapless systems, but exponentially in the presence of a gap.  In the
family of effective models tested here, the correlations are
expected to decay exponentially since the charge sector is gapped.
The single-particle correlations are shown in Fig. \ref{fig:cors}.  In order to compare the
correlations defined in the sliced-basis for the H-chain with those
for the effective model (expressed in the space of atomic sites), we transform the single-particle Green's function of the original
system using the WFs of the corresponding effective model.  Measuring the correlation length in this space gives $l_\text{H-chain}=12.8$ and $l_\text{WF}=13.8$ for $R=2.0a_B$,
decreasing to $l_\text{H-chain}=3.4$ and $l_\text{WF}=3.7$ for $R=3.2a_B$, corresponding to differences of about $8\%$ between the H-chain
and $\hat{H}_\text{WF}$ for both atomic spacings. 
The spin-spin correlations can also be computed and are shown 
in Fig. \ref{fig:sscors}.  

The excellent agreement
of both the correlations 
and spin velocities between $\hat{H}_\text{WF}$ and the H-chain demonstrates the accuracy of 
the procedure used in deriving the WFs and the following
truncation of the 4-index interaction terms.  At this stage,
$\hat{H}_\text{WF}$ constitutes the initial 
set of interactions needed in order to accurately represent the H-chain.  However, one can go further and deduce the minimum set 
of interactions needed to
describe the original system with a certain accuracy.

\begin{figure}
    \centering
    \includegraphics[width=\columnwidth]{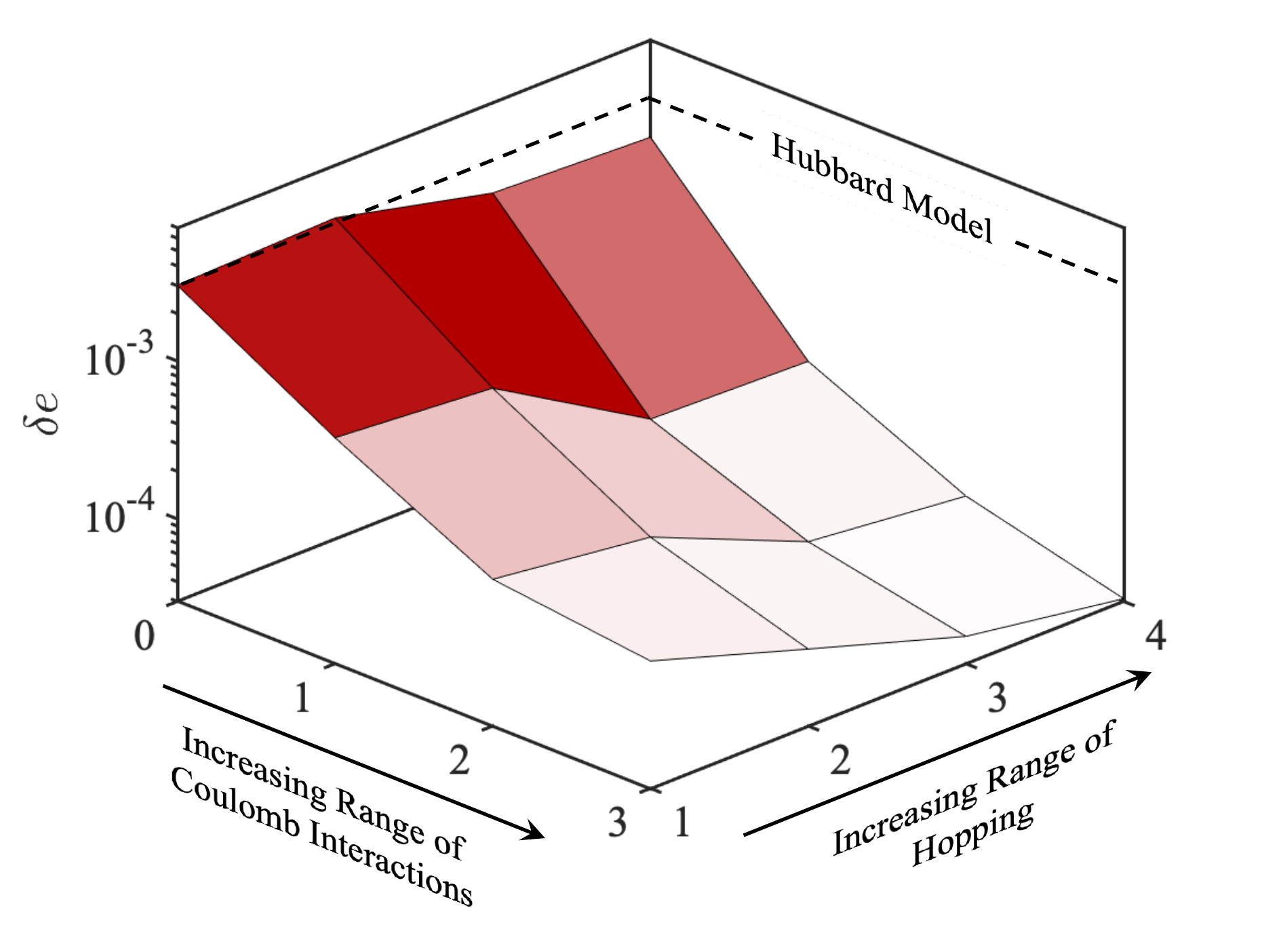}
    \caption{Error for effective models with varying-range one-body and two-body interactions for H20 at $R=2.0a_B$.  
    The dashed black lines correspond to the error made by using only Hubbard interactions
    in the effective model as in (\ref{eqn:Hhub}).  Increasing the range of the two-body interactions
    is seen to be more effective than increasing the range of the one-body hopping.}
    \label{fig:range}
\end{figure}

\paragraph{\textbf{Reducing the Range of Interactions}}

Despite the existence of the full, long-range Coulomb interaction in the original
system, can we reconstruct the ground state and spin excitations  with shorter-range one-body and two-body interaction? Given the 
previously outlined optimization procedure, the importance of these terms can
be tested by fitting to Hamiltonians with increasingly longer-range single-body and
two-body terms.  The performance of each of these models can then be assessed relative to $\hat{H}_\text{WF}$
by calculating,
\begin{align}
    \delta_e = \frac{E_{\text{min}} - E_0^{WF}}{N_a} 
    \label{eqn:opterr}
\end{align}
where $E_{\text{min}}$ is the minimum value resulting from the minimization of  (\ref{eqn:phi}) and $E_0^{WF}$,
the ground state energy of $\hat{H}_\text{WF}$ with $N_a$ atoms.  Small values of $\delta_e$ are indicative
of effective models which are more closely related to $\hat{H}_\text{WF}$. 

\begin{figure}
\centering
\includegraphics[width=\columnwidth]{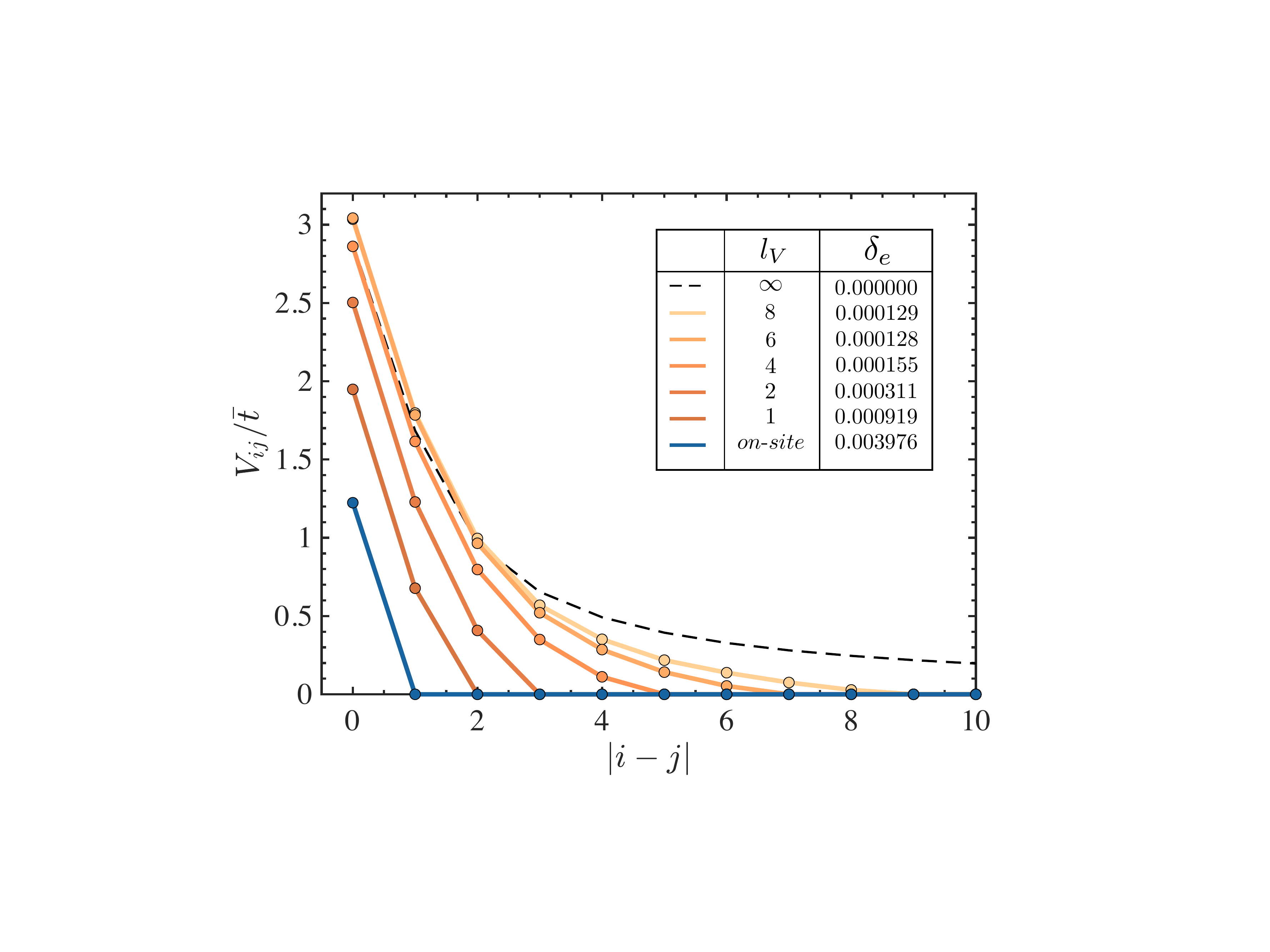}
\caption{Decay of two-body interactions with varying maximum range for H40. The variable $l_V$ is
defined as the maximum range of the two-body interactions.  For example, $l_V = 1$
implies a range up to nearest-neighbor, $l_V=2$, next nearest-neighbor, etc.  The dashed black
line shows the decay of the two-body interactions for $\hat{H}_{WF}$ scaled by 
the average nearest-neighbor hopping.  The other
lines are the two-body interactions for the corresponding effective model with
only a nearest-neighbor hopping term and a limited range two-body interaction.
The error as defined in (\ref{eqn:opterr}) is also shown in the last column of the legend.}
\label{fig:smooth}
\end{figure}


In
Fig. \ref{fig:range} we show the accuracy of different effective models when the range of the hopping and two-body interactions 
are constrained, but with the detailed shape of the interactions optimized subject to the constraint.  For a pure Hubbard model, $\delta_e=3 \times 10^{-3}$, in atomic units.  If we increase only the hopping, the energy improves to $1.7 \times 10^{-3}$ at a range of 3 lattice spacings.  We get a much better improvement if we
keep the hopping nearest-neighbor but increase the Coulomb repulsion out to a range of 3, with the energy error reducing by over an order to magnitude down to $1.9 \times 10^{-4}$. The best improvement comes from increasing both ranges, where we find an energy error well under $10^{-4}$.
The optimal
two-body terms smoothly decay to 
zero with distance as opposed to being sharply cut off.  This decay is shown in
Fig. \ref{fig:smooth} for several effective models with varying maximum interaction
range.  As the maximum range is increased, the two-body interactions smoothly approach
those of the original model.


\begin{figure}
\centering
\includegraphics[width=\columnwidth]{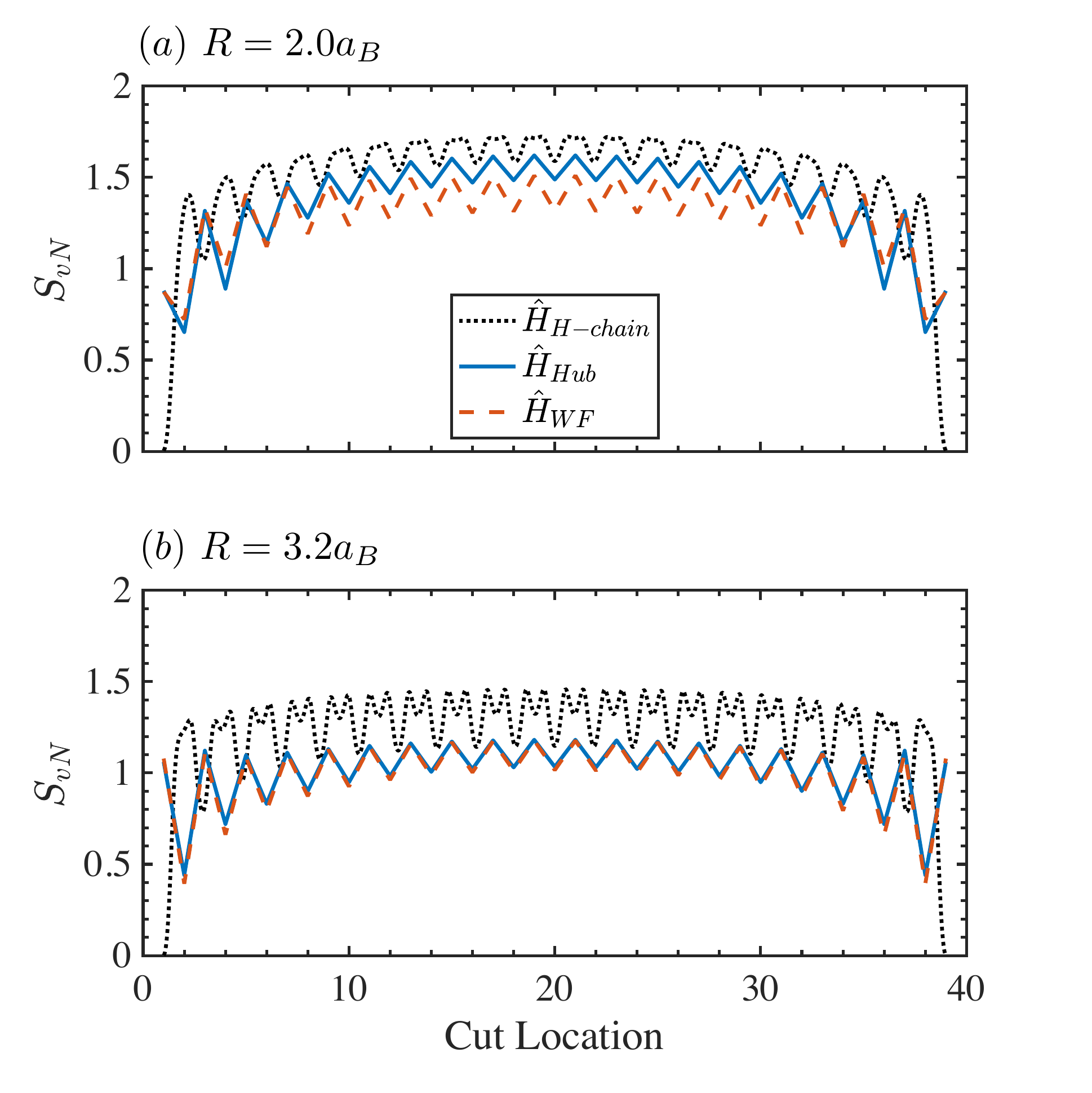}
\caption{The von Neumann entanglement entropy calculated across each bond along the length of chain for $(a) R=2.0a_B$ and $(b) R=3.2a_B$.  The same quantity is plotted for the effective models defined in (\ref{eqn:HWF}) and (\ref{eqn:Hhub}) for comparison.}
\label{fig:entgl}
\end{figure}

It is important to note here that DMRG is based on a low entanglement approximation and that in general, long-range interactions can be a source of high entanglement, hindering its effectiveness.  Indeed there has been an effort to formally prove this, although the proof has not yet been generalized to interactions that decay as $1/r$\cite{KS2020}.  However, in our current H-chain calculations, the Coulomb interaction does not lead to high entanglement and has a similar effect in the H-chain's associated effective models.  Fig. \ref{fig:entgl} compares the entanglement computed across the center of the chain for the H-chain, $\hat{H}_\text{WF}$ and $\hat{H}_\text{Hub}$.  For both $R=2.0a_B$ and $R=3.2a_B$, the entanglement is slightly higher in the H-chain than in either effective model.  One possible source for this higher entanglement could be the short-range dynamical correlations in the sliced basis description, which are not present once one truncates to Wannier functions.  But comparing $\hat{H}_\text{WF}$ to $\hat{H}_\text{Hub}$, we find the entanglement to be lower for $R=2.0a_B$ and nearly identical for $R=3.2a_B$ despite the presence of long-range interactions in $\hat{H}_\text{WF}$.

\begin{figure*}
\centering
\includegraphics[scale=0.5]{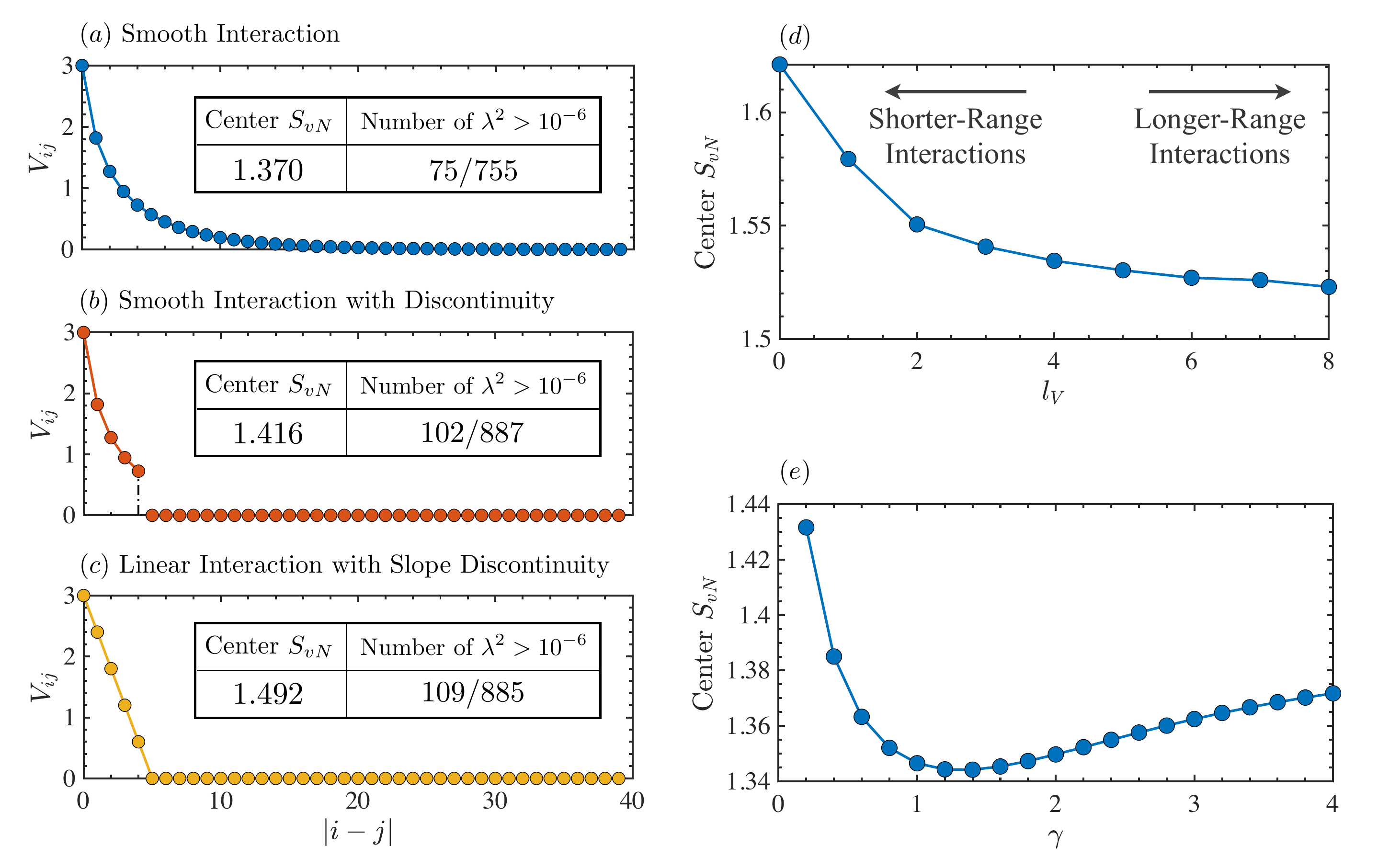}
\caption{A survey of the effects of different interactions on the entanglement entropy.  The models here are all Hubbard models modified with a long-range interaction, with the exception of $(d)$, which uses the optimized models from Fig.\ref{fig:smooth}.  The left column shows the entanglement for $(a)$ a smooth interaction, $(b)$ the same interaction with a discontinuity and $(c)$ the linear approximation to the smooth interaction with a slope discontinuity.  In the right column, the effects of longer-range interactions are shown.  In $(d)$, $l_V$ is defined as the maximum range of the two-body interaction shown in Fig.\ref{fig:smooth} for various ranges.  The entanglement entropy across the center bond for each of these models is plotted.  In $(e)$, the interactions are given by $1/r^\gamma$.  At very long range, the entanglement is large, but then decreases to a minimum as $\gamma \rightarrow 1$.  Then, as the interactions become shorter-range, the entanglement across the center bond is seen to increase.}
\label{fig:LRentgl}
\end{figure*}

A key reason the long range Coulomb interaction does not cause high entanglement could be tied to its smoothness--two particles at large separation interact through changes in the interaction, not the interaction itself, at least in terms of the entanglement of the wavefunction.  Another consideration is the screening of the long-range interaction by the other electrons--a non-smooth interaction may not be as well screened.  We have investigated the relationship between the interaction and the entanglement by studying  a toy extended Hubbard model which allows for unphysical non-smooth long distance terms.  The smooth interaction is shown in Fig. \ref{fig:LRentgl}(a) and its corresponding entanglement in the inset table.  Comparing this to Fig. \ref{fig:LRentgl}(b) and \ref{fig:LRentgl}(c), which features the same interaction with an abrupt discontinuity and its linear approximation with a slope discontinuity, we can see that despite the shorter-range of the latter two interactions, they exhibit higher entanglement.  Similarly, we can examine the entanglement of each of the increasingly longer-range models shown in Fig. \ref{fig:smooth}.  The entanglement as a function of the range of their interactions is plotted in Fig. \ref{fig:LRentgl}(d) which is seen to decrease as the range of the interactions is increased.  Finally, one may attempt to approximate the smooth interaction with a coarser, linearly interpolated interaction hosting much more prominent slope discontinuities.  Fig. \ref{fig:LRentgl}(e) shows the entanglement steadily increasing as the interactions becomes coarser.

The relationship between interaction and entanglement is clearly subtle.  The simple picture that longer-range interactions cause more entanglement is clearly false.  More work is need in this area to clarify how entanglement is altered by long-range interactions with different features.

\begin{figure}
\centering
\includegraphics[width=\columnwidth]{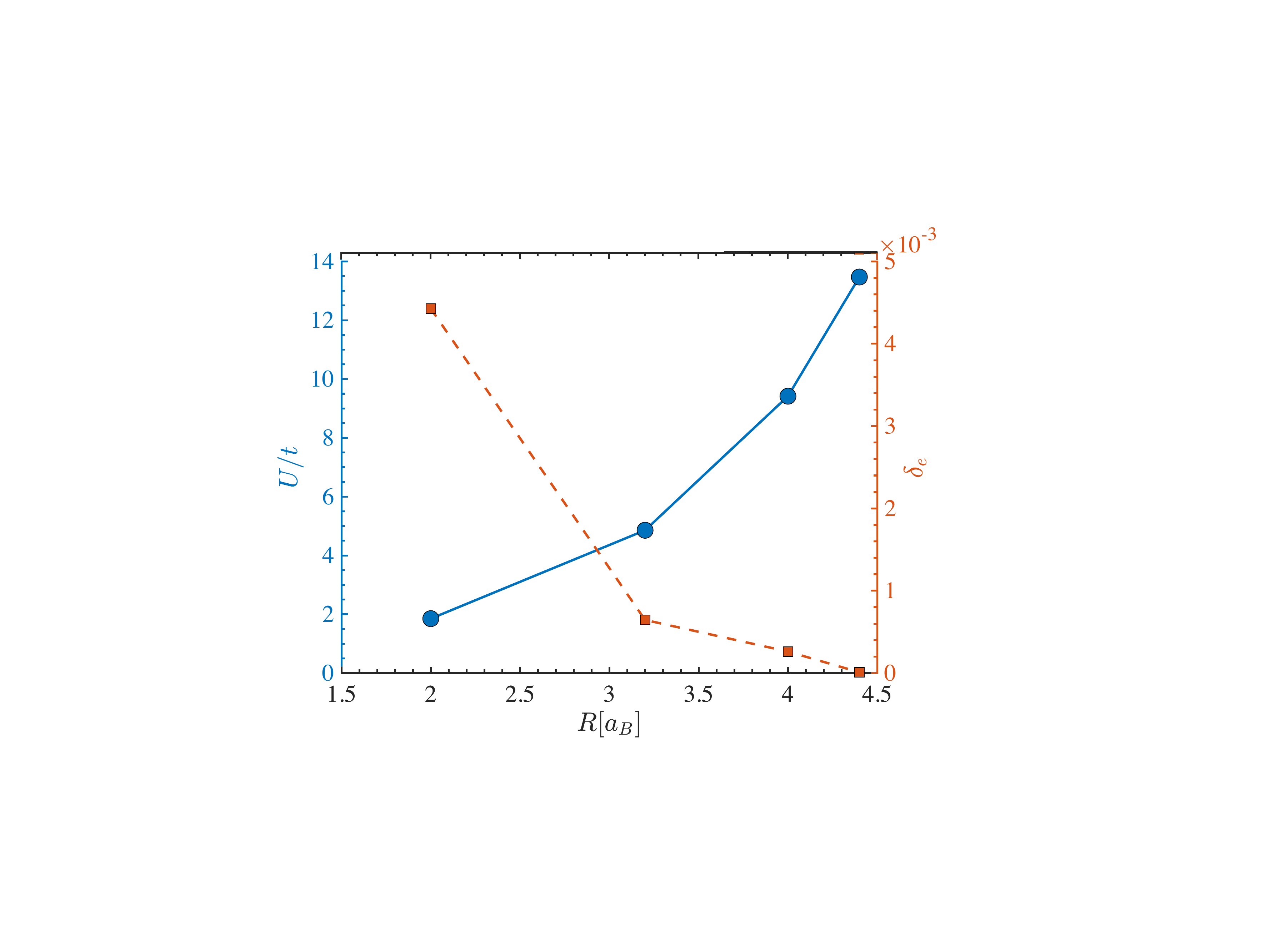}
\caption{Optimal Hubbard parameters for each H-chain as a function of the 
atomic spacing, $R$ for H100.  The resulting Hubbard model becomes ``strongly correlated" near $R=3-4a_B$ where the corresponding value of $U/t$ becomes larger than the bandwidth of about $4t$.  The dashed red line
shows the optimization error as defined in Eq.  (\ref{eqn:opterr}).}
\label{fig:utvr}
\end{figure}

\paragraph{\textbf{Hubbard Model as an Effective Model}}

The range of the one- and two-body interactions can be decreased even further
so that 
the remaining terms in the effective model represent those of a pure 
Hubbard model, with only a nearest-neighbor hopping, $t$, 
an on-site interaction, $U$ and a chemical potential $\mu$.  We allow for slight variations of the model parameters near the edges so that the full Hamiltonian reads
\begin{align}
	\hat{H}_{Hub} = &\sum_{\sigma,\braket{ij}} t^\text{H}_{ij} \cdag{\sigma}{i} \cop{\sigma}{j}\ + 
			\sum_{i} U^\text{H}_i \hat{n}_{\uparrow i} \hat{n}_{\downarrow i} + \nonumber \\
			&\sum_{i} \mu^\text{H}_i \hat{n}_{\text{tot}, i}
	\label{eqn:Hhub}
\end{align}
For small $R$, the effective Hubbard model is 
expected to be in the weakly correlated regime, gradually becoming more and more
correlated as $R$ is increased.  This behavior is shown in Fig. \ref{fig:utvr} for
the optimal Hubbard models.  We can see that beyond $R=3.2a_B$, the effective model
transitions into the strongly-correlated Hubbard regime where it is expected 
to be a better representation for the original H-chain.  This can be seen in
figures \ref{fig:gaps}-\ref{fig:sscors} which
compare the spin velocities and correlations of the H-chains and their 
corresponding effective Hubbard models at $R=2.0a_B$ and $R=3.2a_B$.  

At $R=2.0a_B$, the spin velocities 
for the 1- and 2-spinon pairs in the effective Hubbard model show an 
error of $\sim6\%$ and $\sim8\%$ relative to the original H-chain, 
respectively.  At larger
$R$, these errors decrease to $\sim1\%$ for both exicitations.  Similarly,
the decay of the single-particle Green's function can be measured as the 
slope of the lines shown in Fig. \ref{fig:cors}.  For $R=2.0a_B$, the decay 
rates between the effective Hubbard model and the H-chain differ by $\sim29\%$
whereas at larger $R$, the error decreases to $\sim16\%$.  This main difference
lies within the decay of the tails as an on-site interaction alone can have 
difficulty in replicating the long-range correlations built up from having a 
similarly long-range interaction.  Note that for $\hat{H}_\text{WF}$, the decay 
follows those of the original H-chain much closer in comparison.  This is in 
contrast to the spin-spin correlations which show $\hat{H}_\text{Hub}$ and 
$\hat{H}_\text{WF}$ agreeing almost exactly as $R$ increases from $2.0a_B$ to 
$3.2a_B$.  

The general good agreement between ${H}_\text{Hub}$ and the original H-chain shows that in this
case  the long-range Coulomb interaction is not important in determining the ground state and spin excitations.  However, it is important to remember that this system is at half-filling, and it is insulating. Doping the hydrogen chain is less natural than in a solid:  one would need to add in an
artificial neutralizing charge background. The long range Coulomb interaction is likely more important in the doped case. In addition, we have considered only large enough $R$ so that diffuse bands are not occupied.  The occupation of these outer bands self-dopes the 1S band and induces a metal-insulator transition\cite{Motta_2020}.  In this case the long range Coulomb interaction may  play an important role in both the interactions of the holes in the 1S band and in the physics of the diffuse electrons in the outer bands.


\section{Conclusion}\label{sec:conc}

Using sb-DMRG and the method outlined above, Wannier functions constructed from 
a sum of single-particle density matrices from different spin sectors can be used to 
downfold a chain of hydrogen atoms to a simple effective model. We found that the Wannier
functions constructed from the DMRG natural orbitals are only slightly different from Hartree-Fock Wannier functions, suggesting that for this aspect of downfolding, conventional DFT-based approaches
are likely reliable. These models
capture the key non-charge properties of the low-energy physics of the original system, such as 
spin velocities, single-particle correlations and spin-spin correlations. We were able to construct effective models with just a handful of parameters, with the range of the interactions controlling the
resulting accuracy.  We found that a pure Hubbard model 
with only a nearest-neighbor hopping and on-site interaction
can provide a good representation the ground state and spin-excitation sectors of the H-chain. Further improvements in accuracy are obtained through slightly more extended  interaction terms, reaching
no more that 3 lattice spacings away, provided these interactions are carefully optimized. In this
case the Coulomb terms are found to be more important than the extended hopping terms. In this study
we restricted ourselves to the insulating regime; it is likely that longer-range interactions are needed in the metallic regime of the H-chain.  In the insulating regime, unexpectedly, we find that the entanglement entropy tends to be slightly smaller for models with longer ranged interactions.

\section{Acknowledgements}
RCS was supported by the Simons Foundation through the
Many Electron collaboration. SRW was supported by the
Department of Energy under grant DE-SC0008696.

\bibliographystyle{apsrev4-1}
\bibliography{main}

\begin{thebibliography}{40}%
\makeatletter
\providecommand \@ifxundefined [1]{%
 \@ifx{#1\undefined}
}%
\providecommand \@ifnum [1]{%
 \ifnum #1\expandafter \@firstoftwo
 \else \expandafter \@secondoftwo
 \fi
}%
\providecommand \@ifx [1]{%
 \ifx #1\expandafter \@firstoftwo
 \else \expandafter \@secondoftwo
 \fi
}%
\providecommand \natexlab [1]{#1}%
\providecommand \enquote  [1]{``#1''}%
\providecommand \bibnamefont  [1]{#1}%
\providecommand \bibfnamefont [1]{#1}%
\providecommand \citenamefont [1]{#1}%
\providecommand \href@noop [0]{\@secondoftwo}%
\providecommand \href [0]{\begingroup \@sanitize@url \@href}%
\providecommand \@href[1]{\@@startlink{#1}\@@href}%
\providecommand \@@href[1]{\endgroup#1\@@endlink}%
\providecommand \@sanitize@url [0]{\catcode `\\12\catcode `\$12\catcode
  `\&12\catcode `\#12\catcode `\^12\catcode `\_12\catcode `\%12\relax}%
\providecommand \@@startlink[1]{}%
\providecommand \@@endlink[0]{}%
\providecommand \url  [0]{\begingroup\@sanitize@url \@url }%
\providecommand \@url [1]{\endgroup\@href {#1}{\urlprefix }}%
\providecommand \urlprefix  [0]{URL }%
\providecommand \Eprint [0]{\href }%
\providecommand \doibase [0]{http://dx.doi.org/}%
\providecommand \selectlanguage [0]{\@gobble}%
\providecommand \bibinfo  [0]{\@secondoftwo}%
\providecommand \bibfield  [0]{\@secondoftwo}%
\providecommand \translation [1]{[#1]}%
\providecommand \BibitemOpen [0]{}%
\providecommand \bibitemStop [0]{}%
\providecommand \bibitemNoStop [0]{.\EOS\space}%
\providecommand \EOS [0]{\spacefactor3000\relax}%
\providecommand \BibitemShut  [1]{\csname bibitem#1\endcsname}%
\let\auto@bib@innerbib\@empty
\bibitem [{\citenamefont {Dederichs}\ \emph {et~al.}(1984)\citenamefont
  {Dederichs}, \citenamefont {Bl{\"u}gel}, \citenamefont {Zeller},\ and\
  \citenamefont {Akai}}]{cDFT}%
  \BibitemOpen
  \bibfield  {author} {\bibinfo {author} {\bibfnamefont {P.}~\bibnamefont
  {Dederichs}}, \bibinfo {author} {\bibfnamefont {S.}~\bibnamefont
  {Bl{\"u}gel}}, \bibinfo {author} {\bibfnamefont {R.}~\bibnamefont {Zeller}},
  \ and\ \bibinfo {author} {\bibfnamefont {H.}~\bibnamefont {Akai}},\
  }\href@noop {} {\bibfield  {journal} {\bibinfo  {journal} {Physical review
  letters}\ }\textbf {\bibinfo {volume} {53}},\ \bibinfo {pages} {2512}
  (\bibinfo {year} {1984})}\BibitemShut {NoStop}%
\bibitem [{\citenamefont {Andersen}\ and\ \citenamefont
  {Saha-Dasgupta}(2000)}]{cDFT_application1}%
  \BibitemOpen
  \bibfield  {author} {\bibinfo {author} {\bibfnamefont {O.}~\bibnamefont
  {Andersen}}\ and\ \bibinfo {author} {\bibfnamefont {T.}~\bibnamefont
  {Saha-Dasgupta}},\ }\href@noop {} {\bibfield  {journal} {\bibinfo  {journal}
  {Physical Review B}\ }\textbf {\bibinfo {volume} {62}},\ \bibinfo {pages}
  {R16219} (\bibinfo {year} {2000})}\BibitemShut {NoStop}%
\bibitem [{\citenamefont {Pavarini}\ \emph {et~al.}(2001)\citenamefont
  {Pavarini}, \citenamefont {Dasgupta}, \citenamefont {Saha-Dasgupta},
  \citenamefont {Jepsen},\ and\ \citenamefont {Andersen}}]{cDFT_application2}%
  \BibitemOpen
  \bibfield  {author} {\bibinfo {author} {\bibfnamefont {E.}~\bibnamefont
  {Pavarini}}, \bibinfo {author} {\bibfnamefont {I.}~\bibnamefont {Dasgupta}},
  \bibinfo {author} {\bibfnamefont {T.}~\bibnamefont {Saha-Dasgupta}}, \bibinfo
  {author} {\bibfnamefont {O.}~\bibnamefont {Jepsen}}, \ and\ \bibinfo {author}
  {\bibfnamefont {O.}~\bibnamefont {Andersen}},\ }\href@noop {} {\bibfield
  {journal} {\bibinfo  {journal} {Physical review letters}\ }\textbf {\bibinfo
  {volume} {87}},\ \bibinfo {pages} {047003} (\bibinfo {year}
  {2001})}\BibitemShut {NoStop}%
\bibitem [{\citenamefont {Pines}\ and\ \citenamefont {Bohm}(1952)}]{cRPA1}%
  \BibitemOpen
  \bibfield  {author} {\bibinfo {author} {\bibfnamefont {D.}~\bibnamefont
  {Pines}}\ and\ \bibinfo {author} {\bibfnamefont {D.}~\bibnamefont {Bohm}},\
  }\href@noop {} {\bibfield  {journal} {\bibinfo  {journal} {Physical Review}\
  }\textbf {\bibinfo {volume} {85}},\ \bibinfo {pages} {338} (\bibinfo {year}
  {1952})}\BibitemShut {NoStop}%
\bibitem [{\citenamefont {Bohm}\ and\ \citenamefont {Pines}(1953)}]{cRPA2}%
  \BibitemOpen
  \bibfield  {author} {\bibinfo {author} {\bibfnamefont {D.}~\bibnamefont
  {Bohm}}\ and\ \bibinfo {author} {\bibfnamefont {D.}~\bibnamefont {Pines}},\
  }\href@noop {} {\bibfield  {journal} {\bibinfo  {journal} {Physical Review}\
  }\textbf {\bibinfo {volume} {92}},\ \bibinfo {pages} {609} (\bibinfo {year}
  {1953})}\BibitemShut {NoStop}%
\bibitem [{\citenamefont {Aryasetiawan}\ \emph {et~al.}(2004)\citenamefont
  {Aryasetiawan}, \citenamefont {Imada}, \citenamefont {Georges}, \citenamefont
  {Kotliar}, \citenamefont {Biermann},\ and\ \citenamefont
  {Lichtenstein}}]{Arya2004}%
  \BibitemOpen
  \bibfield  {author} {\bibinfo {author} {\bibfnamefont {F.}~\bibnamefont
  {Aryasetiawan}}, \bibinfo {author} {\bibfnamefont {M.}~\bibnamefont {Imada}},
  \bibinfo {author} {\bibfnamefont {A.}~\bibnamefont {Georges}}, \bibinfo
  {author} {\bibfnamefont {G.}~\bibnamefont {Kotliar}}, \bibinfo {author}
  {\bibfnamefont {S.}~\bibnamefont {Biermann}}, \ and\ \bibinfo {author}
  {\bibfnamefont {A.}~\bibnamefont {Lichtenstein}},\ }\href@noop {} {\bibfield
  {journal} {\bibinfo  {journal} {Physical Review B}\ }\textbf {\bibinfo
  {volume} {70}},\ \bibinfo {pages} {195104} (\bibinfo {year}
  {2004})}\BibitemShut {NoStop}%
\bibitem [{\citenamefont {Belinicher}\ \emph {et~al.}(1994)\citenamefont
  {Belinicher}, \citenamefont {Chernyshev},\ and\ \citenamefont
  {Popovich}}]{Sasha1}%
  \BibitemOpen
  \bibfield  {author} {\bibinfo {author} {\bibfnamefont {V.}~\bibnamefont
  {Belinicher}}, \bibinfo {author} {\bibfnamefont {A.}~\bibnamefont
  {Chernyshev}}, \ and\ \bibinfo {author} {\bibfnamefont {L.}~\bibnamefont
  {Popovich}},\ }\href@noop {} {\bibfield  {journal} {\bibinfo  {journal}
  {Physical Review B}\ }\textbf {\bibinfo {volume} {50}},\ \bibinfo {pages}
  {13768} (\bibinfo {year} {1994})}\BibitemShut {NoStop}%
\bibitem [{\citenamefont {Belinicher}\ and\ \citenamefont
  {Chernyshev}(1994)}]{Sasha2}%
  \BibitemOpen
  \bibfield  {author} {\bibinfo {author} {\bibfnamefont {V.}~\bibnamefont
  {Belinicher}}\ and\ \bibinfo {author} {\bibfnamefont {A.}~\bibnamefont
  {Chernyshev}},\ }\href@noop {} {\bibfield  {journal} {\bibinfo  {journal}
  {Physical Review B}\ }\textbf {\bibinfo {volume} {49}},\ \bibinfo {pages}
  {9746} (\bibinfo {year} {1994})}\BibitemShut {NoStop}%
\bibitem [{\citenamefont {Belinicher}\ \emph
  {et~al.}(1996{\natexlab{a}})\citenamefont {Belinicher}, \citenamefont
  {Chernyshev},\ and\ \citenamefont {Shubin}}]{Sasha3}%
  \BibitemOpen
  \bibfield  {author} {\bibinfo {author} {\bibfnamefont {V.}~\bibnamefont
  {Belinicher}}, \bibinfo {author} {\bibfnamefont {A.}~\bibnamefont
  {Chernyshev}}, \ and\ \bibinfo {author} {\bibfnamefont {V.}~\bibnamefont
  {Shubin}},\ }\href@noop {} {\bibfield  {journal} {\bibinfo  {journal}
  {Physical Review B}\ }\textbf {\bibinfo {volume} {53}},\ \bibinfo {pages}
  {335} (\bibinfo {year} {1996}{\natexlab{a}})}\BibitemShut {NoStop}%
\bibitem [{\citenamefont {Belinicher}\ \emph
  {et~al.}(1996{\natexlab{b}})\citenamefont {Belinicher}, \citenamefont
  {Chernyshev},\ and\ \citenamefont {Shubin}}]{Sasha4}%
  \BibitemOpen
  \bibfield  {author} {\bibinfo {author} {\bibfnamefont {V.}~\bibnamefont
  {Belinicher}}, \bibinfo {author} {\bibfnamefont {A.}~\bibnamefont
  {Chernyshev}}, \ and\ \bibinfo {author} {\bibfnamefont {V.}~\bibnamefont
  {Shubin}},\ }\href@noop {} {\bibfield  {journal} {\bibinfo  {journal}
  {Physical Review B}\ }\textbf {\bibinfo {volume} {54}},\ \bibinfo {pages}
  {14914} (\bibinfo {year} {1996}{\natexlab{b}})}\BibitemShut {NoStop}%
\bibitem [{\citenamefont {Freed}(1983)}]{freed1983}%
  \BibitemOpen
  \bibfield  {author} {\bibinfo {author} {\bibfnamefont {K.~F.}\ \bibnamefont
  {Freed}},\ }\href@noop {} {\bibfield  {journal} {\bibinfo  {journal}
  {Accounts of Chemical Research}\ }\textbf {\bibinfo {volume} {16}},\ \bibinfo
  {pages} {137} (\bibinfo {year} {1983})}\BibitemShut {NoStop}%
\bibitem [{\citenamefont {Zhou}\ and\ \citenamefont
  {Ceperley}(2010)}]{Zhou_2010}%
  \BibitemOpen
  \bibfield  {author} {\bibinfo {author} {\bibfnamefont {S.}~\bibnamefont
  {Zhou}}\ and\ \bibinfo {author} {\bibfnamefont {D.~M.}\ \bibnamefont
  {Ceperley}},\ }\href@noop {} {\bibfield  {journal} {\bibinfo  {journal}
  {Physical Review A}\ }\textbf {\bibinfo {volume} {81}},\ \bibinfo {pages}
  {013402} (\bibinfo {year} {2010})}\BibitemShut {NoStop}%
\bibitem [{\citenamefont {Ten-no}(2013)}]{ten2013}%
  \BibitemOpen
  \bibfield  {author} {\bibinfo {author} {\bibfnamefont {S.}~\bibnamefont
  {Ten-no}},\ }\href@noop {} {\bibfield  {journal} {\bibinfo  {journal} {The
  Journal of chemical physics}\ }\textbf {\bibinfo {volume} {138}},\ \bibinfo
  {pages} {164126} (\bibinfo {year} {2013})}\BibitemShut {NoStop}%
\bibitem [{\citenamefont {G{\l}azek}\ and\ \citenamefont {Wilson}(1993)}]{CT1}%
  \BibitemOpen
  \bibfield  {author} {\bibinfo {author} {\bibfnamefont {S.~D.}\ \bibnamefont
  {G{\l}azek}}\ and\ \bibinfo {author} {\bibfnamefont {K.~G.}\ \bibnamefont
  {Wilson}},\ }\href@noop {} {\bibfield  {journal} {\bibinfo  {journal}
  {Physical Review D}\ }\textbf {\bibinfo {volume} {48}},\ \bibinfo {pages}
  {5863} (\bibinfo {year} {1993})}\BibitemShut {NoStop}%
\bibitem [{\citenamefont {Glazek}\ and\ \citenamefont {Wilson}(1994)}]{CT2}%
  \BibitemOpen
  \bibfield  {author} {\bibinfo {author} {\bibfnamefont {S.~D.}\ \bibnamefont
  {Glazek}}\ and\ \bibinfo {author} {\bibfnamefont {K.~G.}\ \bibnamefont
  {Wilson}},\ }\href@noop {} {\bibfield  {journal} {\bibinfo  {journal}
  {Physical Review D}\ }\textbf {\bibinfo {volume} {49}},\ \bibinfo {pages}
  {4214} (\bibinfo {year} {1994})}\BibitemShut {NoStop}%
\bibitem [{\citenamefont {Wegner}(1994)}]{CT3}%
  \BibitemOpen
  \bibfield  {author} {\bibinfo {author} {\bibfnamefont {F.}~\bibnamefont
  {Wegner}},\ }\href@noop {} {\bibfield  {journal} {\bibinfo  {journal}
  {Annalen der physik}\ }\textbf {\bibinfo {volume} {506}},\ \bibinfo {pages}
  {77} (\bibinfo {year} {1994})}\BibitemShut {NoStop}%
\bibitem [{\citenamefont {White}(2002)}]{CT4}%
  \BibitemOpen
  \bibfield  {author} {\bibinfo {author} {\bibfnamefont {S.~R.}\ \bibnamefont
  {White}},\ }\href@noop {} {\bibfield  {journal} {\bibinfo  {journal} {The
  Journal of chemical physics}\ }\textbf {\bibinfo {volume} {117}},\ \bibinfo
  {pages} {7472} (\bibinfo {year} {2002})}\BibitemShut {NoStop}%
\bibitem [{\citenamefont {Yanai}\ and\ \citenamefont {Chan}(2006)}]{CT5}%
  \BibitemOpen
  \bibfield  {author} {\bibinfo {author} {\bibfnamefont {T.}~\bibnamefont
  {Yanai}}\ and\ \bibinfo {author} {\bibfnamefont {G.~K.-L.}\ \bibnamefont
  {Chan}},\ }\href@noop {} {\bibfield  {journal} {\bibinfo  {journal} {The
  Journal of chemical physics}\ }\textbf {\bibinfo {volume} {124}},\ \bibinfo
  {pages} {194106} (\bibinfo {year} {2006})}\BibitemShut {NoStop}%
\bibitem [{\citenamefont {Acioli}\ and\ \citenamefont
  {Ceperley}(1994)}]{Acioli_1994}%
  \BibitemOpen
  \bibfield  {author} {\bibinfo {author} {\bibfnamefont {P.~H.}\ \bibnamefont
  {Acioli}}\ and\ \bibinfo {author} {\bibfnamefont {D.~M.}\ \bibnamefont
  {Ceperley}},\ }\href@noop {} {\bibfield  {journal} {\bibinfo  {journal} {The
  Journal of chemical physics}\ }\textbf {\bibinfo {volume} {100}},\ \bibinfo
  {pages} {8169} (\bibinfo {year} {1994})}\BibitemShut {NoStop}%
\bibitem [{\citenamefont {Wagner}(2013)}]{Wagner_2013}%
  \BibitemOpen
  \bibfield  {author} {\bibinfo {author} {\bibfnamefont {L.~K.}\ \bibnamefont
  {Wagner}},\ }\href@noop {} {\bibfield  {journal} {\bibinfo  {journal} {The
  Journal of Chemical Physics}\ }\textbf {\bibinfo {volume} {138}},\ \bibinfo
  {pages} {094106} (\bibinfo {year} {2013})}\BibitemShut {NoStop}%
\bibitem [{\citenamefont {Changlani}\ \emph {et~al.}(2013)\citenamefont
  {Changlani}, \citenamefont {Ghosh}, \citenamefont {Pujari},\ and\
  \citenamefont {Henley}}]{Changlani_2013}%
  \BibitemOpen
  \bibfield  {author} {\bibinfo {author} {\bibfnamefont {H.~J.}\ \bibnamefont
  {Changlani}}, \bibinfo {author} {\bibfnamefont {S.}~\bibnamefont {Ghosh}},
  \bibinfo {author} {\bibfnamefont {S.}~\bibnamefont {Pujari}}, \ and\ \bibinfo
  {author} {\bibfnamefont {C.~L.}\ \bibnamefont {Henley}},\ }\href@noop {}
  {\bibfield  {journal} {\bibinfo  {journal} {Physical review letters}\
  }\textbf {\bibinfo {volume} {111}},\ \bibinfo {pages} {157201} (\bibinfo
  {year} {2013})}\BibitemShut {NoStop}%
\bibitem [{\citenamefont {Changlani}\ \emph {et~al.}(2015)\citenamefont
  {Changlani}, \citenamefont {Zheng},\ and\ \citenamefont
  {Wagner}}]{Changlani_2018}%
  \BibitemOpen
  \bibfield  {author} {\bibinfo {author} {\bibfnamefont {H.~J.}\ \bibnamefont
  {Changlani}}, \bibinfo {author} {\bibfnamefont {H.}~\bibnamefont {Zheng}}, \
  and\ \bibinfo {author} {\bibfnamefont {L.~K.}\ \bibnamefont {Wagner}},\
  }\href {\doibase 10.1063/1.4927664} {\bibfield  {journal} {\bibinfo
  {journal} {The Journal of Chemical Physics}\ }\textbf {\bibinfo {volume}
  {143}},\ \bibinfo {pages} {102814} (\bibinfo {year} {2015})},\ \Eprint
  {http://arxiv.org/abs/https://doi.org/10.1063/1.4927664}
  {https://doi.org/10.1063/1.4927664} \BibitemShut {NoStop}%
\bibitem [{\citenamefont {Zheng}\ \emph {et~al.}(2018)\citenamefont {Zheng},
  \citenamefont {Changlani}, \citenamefont {Williams}, \citenamefont
  {Busemeyer},\ and\ \citenamefont {Wagner}}]{zheng2018}%
  \BibitemOpen
  \bibfield  {author} {\bibinfo {author} {\bibfnamefont {H.}~\bibnamefont
  {Zheng}}, \bibinfo {author} {\bibfnamefont {H.~J.}\ \bibnamefont
  {Changlani}}, \bibinfo {author} {\bibfnamefont {K.~T.}\ \bibnamefont
  {Williams}}, \bibinfo {author} {\bibfnamefont {B.}~\bibnamefont {Busemeyer}},
  \ and\ \bibinfo {author} {\bibfnamefont {L.~K.}\ \bibnamefont {Wagner}},\
  }\href@noop {} {\bibfield  {journal} {\bibinfo  {journal} {Frontiers in
  Physics}\ }\textbf {\bibinfo {volume} {6}},\ \bibinfo {pages} {43} (\bibinfo
  {year} {2018})}\BibitemShut {NoStop}%
\bibitem [{\citenamefont {Shinaoka}\ \emph {et~al.}(2015)\citenamefont
  {Shinaoka}, \citenamefont {Troyer},\ and\ \citenamefont
  {Werner}}]{shinaoka2015}%
  \BibitemOpen
  \bibfield  {author} {\bibinfo {author} {\bibfnamefont {H.}~\bibnamefont
  {Shinaoka}}, \bibinfo {author} {\bibfnamefont {M.}~\bibnamefont {Troyer}}, \
  and\ \bibinfo {author} {\bibfnamefont {P.}~\bibnamefont {Werner}},\
  }\href@noop {} {\bibfield  {journal} {\bibinfo  {journal} {Physical Review
  B}\ }\textbf {\bibinfo {volume} {91}},\ \bibinfo {pages} {245156} (\bibinfo
  {year} {2015})}\BibitemShut {NoStop}%
\bibitem [{\citenamefont {Motta}\ \emph {et~al.}(2017)\citenamefont {Motta},
  \citenamefont {Ceperley}, \citenamefont {Chan}, \citenamefont {Gomez},
  \citenamefont {Gull}, \citenamefont {Guo}, \citenamefont {Jim{\'e}nez-Hoyos},
  \citenamefont {Lan}, \citenamefont {Li}, \citenamefont {Ma} \emph
  {et~al.}}]{Motta_2017}%
  \BibitemOpen
  \bibfield  {author} {\bibinfo {author} {\bibfnamefont {M.}~\bibnamefont
  {Motta}}, \bibinfo {author} {\bibfnamefont {D.~M.}\ \bibnamefont {Ceperley}},
  \bibinfo {author} {\bibfnamefont {G.~K.-L.}\ \bibnamefont {Chan}}, \bibinfo
  {author} {\bibfnamefont {J.~A.}\ \bibnamefont {Gomez}}, \bibinfo {author}
  {\bibfnamefont {E.}~\bibnamefont {Gull}}, \bibinfo {author} {\bibfnamefont
  {S.}~\bibnamefont {Guo}}, \bibinfo {author} {\bibfnamefont {C.~A.}\
  \bibnamefont {Jim{\'e}nez-Hoyos}}, \bibinfo {author} {\bibfnamefont {T.~N.}\
  \bibnamefont {Lan}}, \bibinfo {author} {\bibfnamefont {J.}~\bibnamefont
  {Li}}, \bibinfo {author} {\bibfnamefont {F.}~\bibnamefont {Ma}},  \emph
  {et~al.},\ }\href@noop {} {\bibfield  {journal} {\bibinfo  {journal}
  {Physical Review X}\ }\textbf {\bibinfo {volume} {7}},\ \bibinfo {pages}
  {031059} (\bibinfo {year} {2017})}\BibitemShut {NoStop}%
\bibitem [{\citenamefont {Motta}\ \emph {et~al.}(2020)\citenamefont {Motta},
  \citenamefont {Genovese}, \citenamefont {Ma}, \citenamefont {Cui},
  \citenamefont {Sawaya}, \citenamefont {Chan}, \citenamefont {Chepiga},
  \citenamefont {Helms}, \citenamefont {Jim{\'e}nez-Hoyos}, \citenamefont
  {Millis} \emph {et~al.}}]{Motta_2020}%
  \BibitemOpen
  \bibfield  {author} {\bibinfo {author} {\bibfnamefont {M.}~\bibnamefont
  {Motta}}, \bibinfo {author} {\bibfnamefont {C.}~\bibnamefont {Genovese}},
  \bibinfo {author} {\bibfnamefont {F.}~\bibnamefont {Ma}}, \bibinfo {author}
  {\bibfnamefont {Z.-H.}\ \bibnamefont {Cui}}, \bibinfo {author} {\bibfnamefont
  {R.}~\bibnamefont {Sawaya}}, \bibinfo {author} {\bibfnamefont {G.~K.-L.}\
  \bibnamefont {Chan}}, \bibinfo {author} {\bibfnamefont {N.}~\bibnamefont
  {Chepiga}}, \bibinfo {author} {\bibfnamefont {P.}~\bibnamefont {Helms}},
  \bibinfo {author} {\bibfnamefont {C.}~\bibnamefont {Jim{\'e}nez-Hoyos}},
  \bibinfo {author} {\bibfnamefont {A.~J.}\ \bibnamefont {Millis}},  \emph
  {et~al.},\ }\href@noop {} {\bibfield  {journal} {\bibinfo  {journal}
  {Physical Review X}\ }\textbf {\bibinfo {volume} {10}},\ \bibinfo {pages}
  {031058} (\bibinfo {year} {2020})}\BibitemShut {NoStop}%
\bibitem [{\citenamefont {Stoudenmire}\ and\ \citenamefont
  {White}(2017)}]{Stoudenmire_2017}%
  \BibitemOpen
  \bibfield  {author} {\bibinfo {author} {\bibfnamefont {E.~M.}\ \bibnamefont
  {Stoudenmire}}\ and\ \bibinfo {author} {\bibfnamefont {S.~R.}\ \bibnamefont
  {White}},\ }\href {\doibase 10.1103/PhysRevLett.119.046401} {\bibfield
  {journal} {\bibinfo  {journal} {Phys. Rev. Lett.}\ }\textbf {\bibinfo
  {volume} {119}},\ \bibinfo {pages} {046401} (\bibinfo {year}
  {2017})}\BibitemShut {NoStop}%
\bibitem [{\citenamefont {Koch}\ and\ \citenamefont
  {Goedecker}(2001)}]{Koch_Goedecker_2001}%
  \BibitemOpen
  \bibfield  {author} {\bibinfo {author} {\bibfnamefont {E.}~\bibnamefont
  {Koch}}\ and\ \bibinfo {author} {\bibfnamefont {S.}~\bibnamefont
  {Goedecker}},\ }\href {\doibase
  https://doi.org/10.1016/S0038-1098(01)00192-2} {\bibfield  {journal}
  {\bibinfo  {journal} {Solid State Communications}\ }\textbf {\bibinfo
  {volume} {119}},\ \bibinfo {pages} {105 } (\bibinfo {year}
  {2001})}\BibitemShut {NoStop}%
\bibitem [{\citenamefont {Sch{\"u}ler}\ \emph {et~al.}(2013)\citenamefont
  {Sch{\"u}ler}, \citenamefont {R{\"o}sner}, \citenamefont {Wehling},
  \citenamefont {Lichtenstein},\ and\ \citenamefont
  {Katsnelson}}]{Schuler_2018}%
  \BibitemOpen
  \bibfield  {author} {\bibinfo {author} {\bibfnamefont {M.}~\bibnamefont
  {Sch{\"u}ler}}, \bibinfo {author} {\bibfnamefont {M.}~\bibnamefont
  {R{\"o}sner}}, \bibinfo {author} {\bibfnamefont {T.}~\bibnamefont {Wehling}},
  \bibinfo {author} {\bibfnamefont {A.}~\bibnamefont {Lichtenstein}}, \ and\
  \bibinfo {author} {\bibfnamefont {M.}~\bibnamefont {Katsnelson}},\
  }\href@noop {} {\bibfield  {journal} {\bibinfo  {journal} {Physical Review
  Letters}\ }\textbf {\bibinfo {volume} {111}},\ \bibinfo {pages} {036601}
  (\bibinfo {year} {2013})}\BibitemShut {NoStop}%
\bibitem [{\citenamefont {Hachmann}\ \emph {et~al.}(2006)\citenamefont
  {Hachmann}, \citenamefont {Cardoen},\ and\ \citenamefont {Chan}}]{HChain1}%
  \BibitemOpen
  \bibfield  {author} {\bibinfo {author} {\bibfnamefont {J.}~\bibnamefont
  {Hachmann}}, \bibinfo {author} {\bibfnamefont {W.}~\bibnamefont {Cardoen}}, \
  and\ \bibinfo {author} {\bibfnamefont {G.~K.-L.}\ \bibnamefont {Chan}},\
  }\href@noop {} {\bibfield  {journal} {\bibinfo  {journal} {The Journal of
  chemical physics}\ }\textbf {\bibinfo {volume} {125}},\ \bibinfo {pages}
  {144101} (\bibinfo {year} {2006})}\BibitemShut {NoStop}%
\bibitem [{\citenamefont {Tsuchimochi}\ and\ \citenamefont
  {Scuseria}(2009)}]{HChain2}%
  \BibitemOpen
  \bibfield  {author} {\bibinfo {author} {\bibfnamefont {T.}~\bibnamefont
  {Tsuchimochi}}\ and\ \bibinfo {author} {\bibfnamefont {G.~E.}\ \bibnamefont
  {Scuseria}},\ }\href@noop {} {\enquote {\bibinfo {title} {Strong correlations
  via constrained-pairing mean-field theory},}\ } (\bibinfo {year}
  {2009})\BibitemShut {NoStop}%
\bibitem [{\citenamefont {Lin}\ \emph {et~al.}(2011)\citenamefont {Lin},
  \citenamefont {Marianetti}, \citenamefont {Millis},\ and\ \citenamefont
  {Reichman}}]{HChain3}%
  \BibitemOpen
  \bibfield  {author} {\bibinfo {author} {\bibfnamefont {N.}~\bibnamefont
  {Lin}}, \bibinfo {author} {\bibfnamefont {C.}~\bibnamefont {Marianetti}},
  \bibinfo {author} {\bibfnamefont {A.~J.}\ \bibnamefont {Millis}}, \ and\
  \bibinfo {author} {\bibfnamefont {D.~R.}\ \bibnamefont {Reichman}},\
  }\href@noop {} {\bibfield  {journal} {\bibinfo  {journal} {Physical review
  letters}\ }\textbf {\bibinfo {volume} {106}},\ \bibinfo {pages} {096402}
  (\bibinfo {year} {2011})}\BibitemShut {NoStop}%
\bibitem [{\citenamefont {Sinitskiy}\ \emph {et~al.}(2010)\citenamefont
  {Sinitskiy}, \citenamefont {Greenman},\ and\ \citenamefont
  {Mazziotti}}]{HChain4}%
  \BibitemOpen
  \bibfield  {author} {\bibinfo {author} {\bibfnamefont {A.~V.}\ \bibnamefont
  {Sinitskiy}}, \bibinfo {author} {\bibfnamefont {L.}~\bibnamefont {Greenman}},
  \ and\ \bibinfo {author} {\bibfnamefont {D.~A.}\ \bibnamefont {Mazziotti}},\
  }\href@noop {} {\bibfield  {journal} {\bibinfo  {journal} {The Journal of
  chemical physics}\ }\textbf {\bibinfo {volume} {133}},\ \bibinfo {pages}
  {014104} (\bibinfo {year} {2010})}\BibitemShut {NoStop}%
\bibitem [{\citenamefont {Chan}\ \emph {et~al.}(2016)\citenamefont {Chan},
  \citenamefont {Keselman}, \citenamefont {Nakatani}, \citenamefont {Li},\ and\
  \citenamefont {White}}]{ChanWhite2016}%
  \BibitemOpen
  \bibfield  {author} {\bibinfo {author} {\bibfnamefont {G.~K.-L.}\
  \bibnamefont {Chan}}, \bibinfo {author} {\bibfnamefont {A.}~\bibnamefont
  {Keselman}}, \bibinfo {author} {\bibfnamefont {N.}~\bibnamefont {Nakatani}},
  \bibinfo {author} {\bibfnamefont {Z.}~\bibnamefont {Li}}, \ and\ \bibinfo
  {author} {\bibfnamefont {S.~R.}\ \bibnamefont {White}},\ }\href@noop {}
  {\bibfield  {journal} {\bibinfo  {journal} {The Journal of chemical physics}\
  }\textbf {\bibinfo {volume} {145}},\ \bibinfo {pages} {014102} (\bibinfo
  {year} {2016})}\BibitemShut {NoStop}%
\bibitem [{\citenamefont {Parker}\ \emph {et~al.}(2020)\citenamefont {Parker},
  \citenamefont {Cao},\ and\ \citenamefont {Zaletel}}]{Zaletel2020}%
  \BibitemOpen
  \bibfield  {author} {\bibinfo {author} {\bibfnamefont {D.~E.}\ \bibnamefont
  {Parker}}, \bibinfo {author} {\bibfnamefont {X.}~\bibnamefont {Cao}}, \ and\
  \bibinfo {author} {\bibfnamefont {M.~P.}\ \bibnamefont {Zaletel}},\
  }\href@noop {} {\bibfield  {journal} {\bibinfo  {journal} {Physical Review
  B}\ }\textbf {\bibinfo {volume} {102}},\ \bibinfo {pages} {035147} (\bibinfo
  {year} {2020})}\BibitemShut {NoStop}%
\bibitem [{\citenamefont {Marzari}\ and\ \citenamefont
  {Vanderbilt}(1997)}]{marzari1997}%
  \BibitemOpen
  \bibfield  {author} {\bibinfo {author} {\bibfnamefont {N.}~\bibnamefont
  {Marzari}}\ and\ \bibinfo {author} {\bibfnamefont {D.}~\bibnamefont
  {Vanderbilt}},\ }\href@noop {} {\bibfield  {journal} {\bibinfo  {journal}
  {Physical review B}\ }\textbf {\bibinfo {volume} {56}},\ \bibinfo {pages}
  {12847} (\bibinfo {year} {1997})}\BibitemShut {NoStop}%
\bibitem [{\citenamefont {Marzari}\ \emph {et~al.}(2012)\citenamefont
  {Marzari}, \citenamefont {Mostofi}, \citenamefont {Yates}, \citenamefont
  {Souza},\ and\ \citenamefont {Vanderbilt}}]{MLWF}%
  \BibitemOpen
  \bibfield  {author} {\bibinfo {author} {\bibfnamefont {N.}~\bibnamefont
  {Marzari}}, \bibinfo {author} {\bibfnamefont {A.~A.}\ \bibnamefont
  {Mostofi}}, \bibinfo {author} {\bibfnamefont {J.~R.}\ \bibnamefont {Yates}},
  \bibinfo {author} {\bibfnamefont {I.}~\bibnamefont {Souza}}, \ and\ \bibinfo
  {author} {\bibfnamefont {D.}~\bibnamefont {Vanderbilt}},\ }\href {\doibase
  10.1103/RevModPhys.84.1419} {\bibfield  {journal} {\bibinfo  {journal} {Rev.
  Mod. Phys.}\ }\textbf {\bibinfo {volume} {84}},\ \bibinfo {pages} {1419}
  (\bibinfo {year} {2012})}\BibitemShut {NoStop}%
\bibitem [{\citenamefont {Kirtman}(1981)}]{Kirtman1981}%
  \BibitemOpen
  \bibfield  {author} {\bibinfo {author} {\bibfnamefont {B.}~\bibnamefont
  {Kirtman}},\ }\href@noop {} {\bibfield  {journal} {\bibinfo  {journal} {The
  Journal of Chemical Physics}\ }\textbf {\bibinfo {volume} {75}},\ \bibinfo
  {pages} {798} (\bibinfo {year} {1981})}\BibitemShut {NoStop}%
\bibitem [{\citenamefont {Hoffmann}\ and\ \citenamefont
  {Simons}(1988)}]{Hoffman1988}%
  \BibitemOpen
  \bibfield  {author} {\bibinfo {author} {\bibfnamefont {M.~R.}\ \bibnamefont
  {Hoffmann}}\ and\ \bibinfo {author} {\bibfnamefont {J.}~\bibnamefont
  {Simons}},\ }\href@noop {} {\bibfield  {journal} {\bibinfo  {journal} {The
  Journal of chemical physics}\ }\textbf {\bibinfo {volume} {88}},\ \bibinfo
  {pages} {993} (\bibinfo {year} {1988})}\BibitemShut {NoStop}%
\bibitem [{\citenamefont {Kuwahara}\ and\ \citenamefont
  {Saito}(2020)}]{KS2020}%
  \BibitemOpen
  \bibfield  {author} {\bibinfo {author} {\bibfnamefont {T.}~\bibnamefont
  {Kuwahara}}\ and\ \bibinfo {author} {\bibfnamefont {K.}~\bibnamefont
  {Saito}},\ }\href@noop {} {\bibfield  {journal} {\bibinfo  {journal} {Nature
  communications}\ }\textbf {\bibinfo {volume} {11}},\ \bibinfo {pages} {1}
  (\bibinfo {year} {2020})}\BibitemShut {NoStop}%
\end{thebibliography}%

\end{document}